%% file: main.tex
\documentclass{aa}  

\usepackage{graphicx}
\usepackage{txfonts}
\usepackage{hyperref}
\usepackage{newtxtext}
\usepackage[varvw]{newtxmath}

\usepackage{amssymb}
\usepackage{lipsum}
\usepackage{afterpage}
\usepackage{threeparttable}
\usepackage{placeins}
\usepackage{multirow,makecell,booktabs}
\usepackage{graphicx}
\usepackage{subfig}
\usepackage[nameinlink]{cleveref}

\Crefname{figure}{Fig.}{Figs.}
\Crefname{equation}{Eq.}{Eqs.}
\Crefname{section}{Sect.}{Sects.}
\usepackage{enumitem}

\newcommand{\hi}{\textsc{Hi}}

\newcommand{\ha}{H$\alpha$}

\newcommand{\defhi}{$\rm def_\textsc{Hi}$}

\newcommand{\Mo}{M_{\odot}}

\newcommand{\Mhi}{{M_{\textsc{Hi}}}}

\newcommand{\cm}{\rm cm^{-2}}
\newcommand{\kms}{km\,s$^{-1}$}

\newcommand{\e}[1]{\times 10^{#1}}

\newcommand{\sofia}{\textsc{SoFiA}}

\hypersetup{
    colorlinks = true,
    citecolor = blue,
    linkcolor = blue
}

\begin{document} 

\title{MeerKAT view of Hickson Compact Groups:}
\subtitle{II. \hi\ deficiency in the core and surrounding regions}
\titlerunning{\hi\ deficiency in HCGs}
\authorrunning{Sorgho et al.}

\author{A. Sorgho\inst{1},
      L. Verdes-Montenegro\inst{1},
      R. Ianjamasimanana\inst{1},
      K. M. Hess\inst{2,3,1},
      M. G. Jones\inst{4},
      M. Korsaga\inst{5,1},
      Jing Wang\inst{6},
      Xuchen Lin\inst{7},
      J. M. Solanes\inst{8,9},
      M. E. Cluver\inst{10},
      J. M. Cannon\inst{11},
      A. Bosma\inst{12},
      E. Athanassoula\inst{12},
      A. del Olmo\inst{1},
      J. Perea\inst{1},
      J. Mold\'on\inst{1},
      T. Wiegert\inst{1},
      S. Sanchez-Exp\'osito\inst{1},
      J. Garrido\inst{1},
      R. Garc\'ia-Benito\inst{1},
      G. I. G. J\'ozsa\inst{13,14},
      S. Borthakur\inst{15},
      T. Jarrett\inst{16,17},
      B. Namumba\inst{1},
      E. P\'erez\inst{1},
      Javier Rom\'an\inst{18},
      O. M. Smirnov\inst{19,20,21},
      M. Yun\inst{22}
      }

\institute{\inst{1} Instituto de Astrof\'isica de Andaluc\'ia (CSIC), Glorieta de la Astronom\'ia s/n, 18008 Granada, Spain\\
          \email{asorgho@iaa.es}\\
    \inst{2} Department of Space, Earth and Environment, Chalmers University of Technology, Onsala Space Observatory, 43992 Onsala, Sweden\\
    \inst{3} Netherlands Institute for Radio Astronomy (ASTRON), Postbus 2, 7990 AA Dwingeloo, the Netherlands\\
    \inst{4} Steward Observatory, University of Arizona, 933 North Cherry Avenue, Rm. N204, Tucson, AZ 85721-0065, USA\\
    \inst{5} Laboratoire de Physique et de Chimie de l'Environnement, Universit\'e Joseph Ki-Zerbo, 03 BP 7021, Ouaga 03, Burkina Faso\\
    \inst{6} Kavli Institute for Astronomy and Astrophysics, Peking University, Beijing 100871, People’s Republic of China\\
    \inst{7} Department of Astronomy, School of Physics, Peking University, Beijing 100871, People’s Republic of China\\
    \inst{8} Departament de F\'isica Qu\`antica i Astrof\'isica, Universitat de Barcelona, C. Mart\'i i Franqu\`es 1, 08028, Barcelona, Spain\\
    \inst{9} Institut de Ci\`encies del Cosmos (ICCUB), Universitat de Barcelona., C. Mart\'i i Franqu\`es 1, 08028, Barcelona, Spain\\
    \inst{10} Centre for Astrophysics and Supercomputing, Swinburne University of Technology, Hawthorn, VIC 3122, Australia\\
    \inst{11} Department of Physics \& Astronomy, Macalester College, 1600 Grand Avenue, Saint Paul, MN 55105, USA\\
    \inst{12} Aix Marseille Univ, CNRS, CNES, LAM, Marseille, France\\
    \inst{13} Max-Planck-Intitut f\"ur Radioastronomie, Auf dem H\"ugel 69, 53121 Bonn, Germany\\
    \inst{14} Department of Physics and Electronics, Rhodes University, PO Box 94, Makhanda 6140, South Africa\\
    \inst{15} School of Earth and Space Exploration, Arizona State University, 781 Terrace Mall, Tempe, AZ, 85287, USA\\
    \inst{16} Astronomy Department, University of Cape Town, Private Bag X3, Rondebosch 7701, South Africa\\
    \inst{17} Inter-University Institute for Data Intensive Astronomy (IDIA), University of Cape Town, Rondebosch, Cape Town, 7701, South Africa\\
    \inst{18} Departamento de F\'isica de la Tierra y Astrof\'isica, Universidad Complutense de Madrid, E-28040 Madrid, Spain\\
    \inst{19} Centre for Radio Astronomy Techniques and Technologies (RATT), Department of Physics and Electronics, Rhodes University, Makhanda, 6140, South Africa\\
    \inst{20} South African Radio Astronomy Observatory, Black River Park, 2 Fir Street, Observatory, Cape Town 7925, South Africa\\
    \inst{21} Institute for Radioastronomy, National Institute of Astrophysics (INAF IRA), Via Gobetti 101, 40129 Bologna, Italy\\
    \inst{22} Department of Astronomy, University of Massachusetts, Amherst, MA 01003, USA
         }

\date{Received xxx / Accepted xxx}

\abstract
{Hickson compact groups (HCGs) offer an ideal environment for investigating galaxy transformation as a result of interactions. It has been established that the evolutionary sequence of HCGs is marked by an intermediate stage characterised by a substantial amount of \hi\ in their intragroup medium (IGrM) in the form of tidal tails and bridges (Phase 2), rapidly followed by a final stage where no IGrM gas is found and where their member galaxies are highly \hi\ deficient (Phase 3).}
{Despite numerous single-dish and interferometric \hi\ studies on the HCGs, a clear \hi\ picture of the groups within their large-scale environment still remains to be uncovered. Taking advantage of the MeerKAT's high column density sensitivity and large field-of-view, we aim to investigate the rapid transformation of HCGs from the intermediate to late phases, and establish a picture of their gas content variations in the context of their large-scale environments.}
{We performed MeerKAT observations of six HCGs selected to represent the intermediate and late phases of the proposed evolutionary sequence. Combining the \hi\ observations to data from recent wide-field optical surveys, we evaluated the \hi\ deficiencies of galaxies in a ${\sim}30'$ radius of the HCGs.}
{We find that galaxies surrounding both phases exhibit similar distributions in their gas content. Similarly, galaxies making up the cores of Phase 2 HCGs are comparable to their neighbours in terms of \hi\ deficiencies. However, Phase 3 groups are over an order of magnitude more deficient than their surroundings, supporting previous findings that late-phase HCG galaxies are more evolved than their large-scale environments.}
{}


\maketitle

\section{Introduction}
Standard models of structure formation predict that individual galaxies assemble hierarchically to form larger structures such as groups and clusters in the Universe \citep[e.g.,][]{White1978,Blumenthal1984,Springel2005}. Observations of galaxies in environments with different densities have revealed a systematic difference in morphological distributions, with dense environments predominantly populated by early-type, gas-deficient galaxies while the late-type, gas-rich galaxies tend to reside in low-density environments \citep{Dressler1980,Goto2003}. It is believed that galaxies undergo gas-stripping mechanisms when they fall within high-density structures, such as clusters \citep{Gunn1972,Cayatte1990,Vollmer2001,Ramatsoku2019,Ramatsoku2020,Moretti2023}. More recent studies have shown that a significant part of these gas-removal processes, a phase known as pre-processing, takes place inside galaxy groups \citep[e.g.,][]{Hess2013,Bahe2013,Jones2020}. The most well-studies cases of gas stripping processes include substructures within the Virgo cluster \citep[e.g.,][]{Kenney2004,Chung2007,Sorgho2017} and within the Hydra cluster \citep[e.g.][]{Hess2022}. Similar processes have also been observed in group environments within other nearby clusters \citep[e.g.,][]{Dzudzar2019a,Healy2021} where the effects of these mechanisms on the \hi\ content have been demonstrated. This makes galaxy groups an ideal place to investigate processes affecting galaxy properties that may lead to their transformation throughout their lifetime.

A type of galaxy aggregation that deserves particular interest is compact groups, namely the Hickson Compact Groups \citep{Hickson1982}. These are collections of four to ten galaxies distributed in compact configurations and located in low-density environments, often showing signs of extreme gravitational interactions. They present low velocity dispersions, typically ${\lesssim}200\rm\,km\,s^{-1}$ \citep{Hickson1992}. Due to their compactness, HCGs offer unique laboratories for studying the effects of repeated tidal encounters and the role of intragroup gas on galaxy evolution.

Various \hi\ studies have found that several of the catalogued HCGs are \hi\ deficient \citep{Huchtmeier1997,Verdes2001,Williams2002}, presumably because they have lost a significant fraction of their cold gas reservoir throughout their evolution. In the first large study of the \hi\ content of HCGs, \citet{Verdes2001} studied 72 HCGs, of which 16 were imaged at high resolution with the Very Large Array (VLA) telescope and the remaining with single-dish telescopes. The authors proposed an evolutionary sequence of HCGs with four phases: Phase 1 where the \hi\ distribution and kinematics are unperturbed and ${\gtrsim}90\%$ of gas is in the discs of the galaxy members, Phase 2 where 30\% to 60\% of the gas is in the form of tidal features, and Phase 3a in which most of the \hi\ is stripped from the disc and is either found in tails or altogether missing. Furthermore, an additional and uncommon Phase 3b consists of a stage where the essential part of the \hi\ in the group core is contained in large \hi\ clouds with a single velocity gradient. Later, \citet{Borthakur2010} performed deep single-dish GBT observations of 22 HCGs covering all three main phases, measuring an average mass of $8\times10^9\rm\,\Mo$ per group. In a comparative study with interferometric VLA data compiled from various sources, the authors recovered excess \hi\ (with respect to VLA-measured masses) in all HCGs. They particularly found that the highest fraction of excess gas was observed in Phase 2 groups, with an average excess-to-measured-mass fraction of 48\%. The most recent study on the subject, \citet{Jones2023}, followed up on these investigations by reprocessing VLA data of 38 HCGs and found that Phases 1 and 2 have consistent \hi\ deficiencies. However, Phase 3 groups are found to be much more deficient, with 90\% of their gas missing. They also identified a new Phase 3c characterised by a single \hi-bearing galaxy in an otherwise highly \hi-deficient group, likely the result of an evolved group acquiring a new gas-rich member. However, due to the limited column density sensitivity of interferometers, it is unclear whether the cores of the late-phase HCGs studied in \citet{Jones2023} are indeed \hi-deficient or, in fact, contain an amount of \hi\ in a diffuse form that is not revealed by these observations. 
The environmental mechanisms responsible for gas depletion in group galaxies can be categorized into two types: tidal interactions between a galaxy and its neighbours, and interactions with the intragroup medium (IGrM). While evidence is found for both types of interactions in HCGs, several studies reveal that a significant fraction of compact groups lack prominent X-ray emission, a tracer of the hot gas in the IGrM. In those groups, either no diffuse hot gas is detected, or it is linked to individual galaxies \citep[e.g.,][]{Rasmussen2008,Desjardins2013}. Furthermore, \citet{Desjardins2014} found that high luminosity X-ray emission tends to be associated with high baryonic mass compact groups, while those of lower masses are observed to host localised X-ray emission. The study further found that the most X-ray luminous groups experience the lowest star formation rates (SFRs), argued to be the consequence of the lack of cold gas to form stars. However, the X-ray luminosity is not correlated with the \hi\ deficiency of the groups, such that not all gas-deficient HCGs are X-ray luminous and vice-versa. This favours the argument that galaxy-IGrM interactions are not the dominant processes responsible for \hi\ removal from galaxies in the cores of HCGs \citep{Rasmussen2008}, although they remain important \citep[e.g.,][]{Deb2023}. This highlights the important role of tidal interactions in the observed \hi\ deficiency of HCG cores. It is believed that these galaxy-galaxy interactions, more frequent in the past, have transformed the cores of the groups into evolved systems that are embedded into more extended structures evolving more slowly \citep{Coziol1998}. This picture is supported by the more recent analysis of the molecular gas in galaxies selected in 12 HCGs, revealing that the most evolved galaxies in these systems experience the most severe star formation suppression, caused by a disruption of the existing molecular gas in the group \citep{Alatalo2015}. 

Thanks to the upgraded capabilities of the MeerKAT telescope, a more complete \hi\ map of six HCGs selected in phases 2 and 3 has recently become available \citep{Ianja2025}. In this work, we present a detailed \hi\ deficiency study of galaxy members of these HCG groups. In particular, we examine the distribution of the deficiency in the groups and their immediate surrounding, and investigate any link between the groups and their environment in terms of gas depletion. Through this study, we aim to understand how HCGs transition from an impressive complex of \hi\ tidal structures and diffuse gas to an evolved state in which galaxies have completely lost their \hi. Furthermore, we aim to uncover the full picture of the effects of the rapid transformation on the gas content of galaxies in the larger scale HCGs environments.

The paper is organised as follows: in \Cref{sec:sample} we describe the sample and the \hi\ observations; in \Cref{sec:optical} we detail the procedures followed to measure the optical photometry of the galaxies in the HCG environments and determine their stellar masses. Next, we present a census of the \hi\ in the HCGs in \Cref{sec:census} and evaluate their deficiencies in \Cref{sec:hidef}. Finally, we discuss the observed \hi\ deficiencies as well as their variations in \Cref{sec:discussion} and summarise the results in \Cref{sec:summary}. Throughout the paper, we adopt a Hubble constant of $H_0 = 70~\rm km\,s^{-1}\,Mpc^{-1}$.

\section{Sample, \hi\ observations and data}\label{sec:sample}
We have observed six Hickson compact groups, with at least four spectroscopically confirmed members, with the MeerKAT telescope. The observations, conducted between July 2021 and January 2022, made use of the telescope's $L$-band and 32k correlator with a spectral resolution of ${\sim}5.5\rm\,km\,s^{-1}$. Furthermore, to maximise the column density sensitivity, we ensured that at least 61 of the array's 64 antennas were available during the observations. 

The target HCG groups were selected to equally represent the main intermediate and late phases (i.e, Phases 2 \& 3) of the evolutionary sequence (\Cref{tab:hcg_props}). Processes responsible for depleting, dispersing, or consuming the gas are most active in Phase 2, while Phase 3 groups represent the most extreme stages of evolution, raising the most questions about the fate of the \hi. We selected three groups in each phase to cover these processes, and to build a cohesive picture of the rapid evolution of HCGs using the new findings we expect to obtain. An extensive description of the observations, as well as of the data reduction, is given in \citet{Ianja2025}. For each group, several datacubes of varying resolutions were produced to highlight scales of the IGrM. However, for the purpose of this work, we only consider the arcminute-resolution cube\footnote{Except for HCG 31 where we also use the $15''$-resolution datacube for the disc separation, as noted in \Cref{sec:census:sep}.}, which represents the best compromise between column density sensitivity and resolution.

\input{tables/hcg_props.tex}

The HCGs were observed down to a $3\sigma$ column density sensitivity of ${\sim}3.2-3.7\e{18}\rm\,cm^{-2}$ over 20 \kms, at a restoring beam of about an arcminute (translating to a physical scale of ${\sim}9{-}26$~kpc, see \Cref{tab:hcg_obs}). For each HCG, the \hi\ mass limit corresponding to the $3\sigma$ noise level at the centre of the group is given in the table (see \Cref{sec:census:mass} for details).

\begin{table*}[]
    \centering
    \small
    \caption{Parameters of the HCG \hi\ cubes observed with MeerKAT.}
    \label{tab:hcg_obs}
    \begin{tabular}{c c c c c c c}
    \hline \hline
    \multirow{2}{*}{Phase} & \multirow{2}{*}{HCG} & $\theta_{\rm maj}\times\theta_{\rm min}$ & $\theta_{\rm maj}\times\theta_{\rm min}$ & $\rm noise_{3\sigma}$ & $\rm N_{\hi,3\sigma}$ & $\rm M_{\hi,{\rm limit}}$\\
    & & $\rm (arcsec^2)$ & $\rm (kpc^2)$ & ($\rm mJy~beam^{-1}$) & ($10^{18}\,\rm cm^{-2}$) & ($\rm \log\,M_\odot$)\\
    (1) & (2) & (3) & (4) & (5) & (6) & (7) \\
    \hline \rule{0pt}{10pt}
    \multirow{3}{*}{2} & 16 & 57.9 $\times$ 57.5 & $13.8\times13.3$ & 0.33 & 3.5 & 7.3 \\
    & 31  & 59.9 $\times$ 59.2 & $15.4\times14.4$ & 0.33 & 3.3 & 7.4 \\
    & 91  & 58.0 $\times$ 56.1 & $25.9\times25.0$ & 0.33 & 3.7 & 7.9 \\
    \hline \rule{0pt}{10pt}
    \multirow{3}{*}{3} & 30  & 60.5 $\times$ 59.8 & $17.1\times16.6$ & 0.33 & 3.3 & 7.4 \\
    & 90  & 57.7 $\times$ 56.7 & $9.3\times9.0$ & 0.32 & 3.4 & 7.0 \\
    & 97  & 60.2 $\times$ 59.3 & $23.9\times23.1$ & 0.32 & 3.2 & 7.7 \\
    \hline
    \end{tabular}
\tablefoot{
The columns are: (1) the HCG phase; (2) the HCG group ID; (3) the synthesised beam of the considered datacube; (4) the physical scale corresponding to the synthesised beam; (5) the $3\sigma$ noise level in the datacube per 5.5~\kms\ channel; (6) the column density corresponding to the $3\sigma$ noise level over a 20~\kms\ linewidth; (7) the \hi\ mass limit corresponding to the $3\sigma$ noise level.}
\end{table*}


\section{Optical properties}\label{sec:optical}

\subsection{Optical photometries}\label{sec:optical:phot}
Partially because of their location in the southern hemisphere, few galaxies in the vicinity of the selected HCGs are included in existing photometric catalogues. Here, we present new photometric measurements for all galaxies within a $30'$ radius of the HCGs. Five of the six HCGs fall within the footprint of the DECam Legacy Survey \citep[DECaLS;][]{Dey2019} DR10\footnote{\url{https://www.legacysurvey.org/dr10}}. The survey covers the entire South Galactic Cap and the region of the North Galactic Cap at declinations below $34^\circ$, reaching an $r$-band surface brightness limit of ${\sim}27.9\rm\,mag\,arcsec^{-2}$ in the regions of the HCGs. For each of the five HCGs, we have retrieved a ${\sim}1^\circ\times1^\circ$ cutout image in the $g$ and $r$ bands, centred on the group centre. For the last group, HCG 91, only $g$- and $i$-band images are available; we discuss this at the end of this section.

To perform the photometric measurements from the DECaLS images, we made $6'\times6'$ cutouts (enough to cover the spatial extents of the galaxies) centred on each galaxy. We then adapted the procedure described in \citet{Wang2017} and further improved in \citet{Lin2023}, which mainly consists of (i) masking the images, (ii) subtracting the background, (iii) constructing a brightness profile and finally (iv) measuring the total flux. In the masking step, we considered the $r$-band images in order to produce mask files with the same pixel sizes as the cutout images. This consisted of masking emissions from all neighbouring sources using the {\sc Photutils} package \citep{Bradley2024}. We further deblended interacting sources, mainly in the group cores,  to separate the emission of the galaxy of interest from that of the interacting counterparts. The output mask is a combination of two masks, each respectively produced using parameters corresponding to {\sc SExtractor}'s ``cold'' and ``hot'' modes as described in \citet{Rix2004}. While, in the ``cold'' mode, we focus on carefully separating the sources from their neighbours, in the ``hot'' mode, however, the effort is put into finding and masking stars located away from the galaxy centre.

Next, we performed background subtraction by modelling the image background with a two-dimensional equation. This is then subtracted from the image, and the masked pixels were replaced with the averaged pixel value around the galaxy centre. In the third step, concentric annuli are generated using the geometric parameters obtained from the $r$-band images. The surface brightness is measured as the $3\sigma$ clipped average value in each of these annuli. However, a condition is set in the outer region of the galaxy to ensure the flattening of the profile: the surface brightness values of 15 contiguous annuli must be linearly uncorrelated with their sizes. When this condition is met, a $3\sigma$ clipped average of those 15 values is subtracted as residue background.

Lastly, to estimate the total flux, we simply measure the growth curve of the flux within each annulus as a function of aperture size. Similarly to the surface brightness, the flattening condition must be met by the flux values of the individual annuli. When this is the case, the total flux of the galaxy is obtained from the $3\sigma$ clipped average of the fluxes of the outer annuli. This is performed for each of the $g$- and $r$-band images to obtain their respective total fluxes, which are then converted into magnitudes.

For the last group, HCG 91, we retrieved the $r$- and $g$-band magnitudes of galaxies within a $30'$ radius of the group centre from the photometric catalogues of the fourth Data Release\footnote{\url{https://splus.cloud/documentation/DR4}} of the Southern Photometric Local Universe Survey \citep[S-PLUS;][]{DeOliveira2019}. The S-PLUS observations are performed on the robotic telescope T80-South, located at the Cerro Tololo Observatory. The DR4 of S-PLUS includes value-added catalogues, in which star-galaxy-quasar classification and photometric redshifts are provided for millions of objects \citep{Herpich2024}. We proceeded to select all objects falling within the region of HCG 91, detected with a signal-to-noise ratio S/N{>}3 and classified as a galaxy with a probability ${\geq}90\%$. Additionally, to discard background sources, we selected objects with photometric redshift ${\lesssim}0.03$ (the approximate upper limit for the group's redshift) and required that the redshift measurement be reliable. The photometric redshift of the S-PLUS detections is estimated through a supervised machine learning algorithm, and the value quoted for a given source in the value-added catalogue represents the maximum-a-posteriori estimate from the probability density function (PDF) of the source \citep{Herpich2024}. We define a redshift estimate as reliable if its associated error (the difference between the 16th and 84th percentile values of the PDF) is less than 0.02. For context, this error represents the scatter of the S-PLUS redshift measurements \citep{Lima2022}. Lastly, for a source to be selected, for each candidate galaxy, we check in NED\footnote{NASA/IPAC Extragalactic Database, \url{https://ned.ipac.caltech.edu}} and the DECaLS $g$ images for confirmation and search for available properties such as the morphology. A total of 31 galaxies were confirmed in the region of HCG 91 (including the four members of the group), of which we retrieved the $g$ and $r$ magnitudes from the S-PLUS catalogue. 

Finally, we corrected all magnitudes for Galactic extinction based on the dust map of \citet{Schlegel1998}.

\subsection{Stellar masses}\label{sec:optical:smass}
\begin{figure}
    \centering
    \includegraphics[width=\columnwidth]{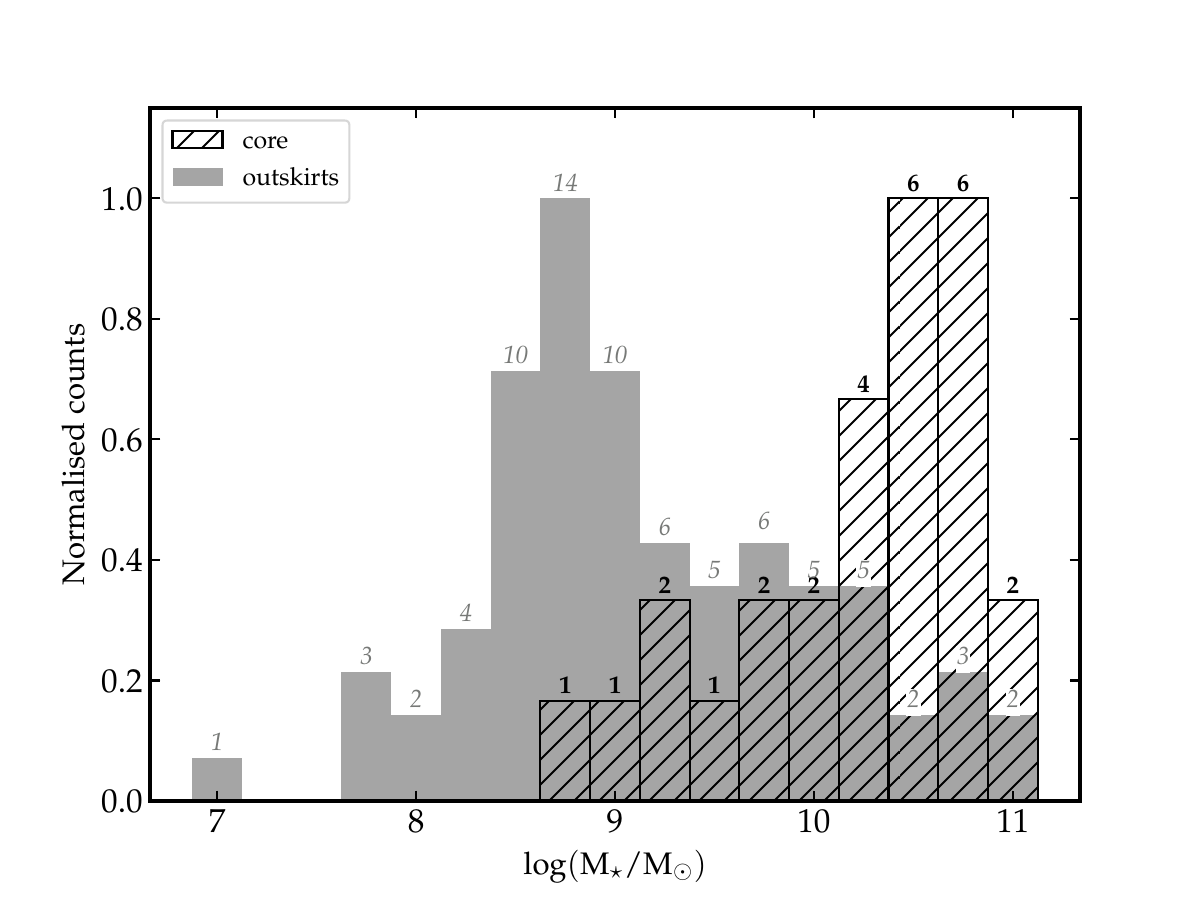}
    \vspace{-20pt}
    \caption{Stellar mass distribution in the HCGs cores (hatched bars) and their surrounding galaxies (solid bars). For visibility, the histograms are normalised by the maximum of each population. The number above each bar represents the absolute count in the corresponding bin.}
    \label{fig:dist_smass}
\end{figure}

To estimate the stellar masses of the galaxies, we take advantage of their intrinsic relation with the galaxy colour. Thus, by considering the $r$-band absolute magnitude $M_r$ and the $g-r$ colour of the galaxies, we use the equation:
\begin{equation}
    \log{(M_{\star}/M_\odot)} = a_r (g-r) + b_r - 0.4({\rm M}_r-{\rm M}_{r,\odot})
\end{equation}
where ${\rm M}_{r,\odot} = 4.65$ is the Sun's absolute magnitude in the $r$-band \citep{Willmer2018} and $a_r=1.49$ and $b_r=-0.70$ are obtained in \citet{GarciaBenito2019} for a sample of galaxies including various morphological types and spanning a stellar mass range of $10^{8.4}{-}10^{12}\,\Mo$. The distribution of the stellar masses in the vicinity of the HCGs is presented in \Cref{fig:dist_smass}. The most massive galaxies in the sample are located in the group cores, with the surrounding galaxies having a median stellar mass of 1.2 dex lower.

\subsection{HCG centres}\label{sec:optical:centres}
The centre coordinates of the HCGs in the literature are from various sources, employing different techniques and considerations to determine the centres of the groups. For example, the central coordinates of HCG 16 on NED were computed by \citet{Barton1996}, whose group definition criteria differ from those of \citet{Hickson1982}. For a more consistent analysis, we adopt a homogeneous method for estimating the centre of the HCGs based on the positions and stellar masses of their core members.
We include in the core all galaxies previously identified as members of the groups, consistently with \citealt{Jones2023} and earlier studies. These members were first selected such that the groups contain at least four members within a 3-magnitude range of the brightest galaxy. Furthermore, to ensure compactness and isolation, a limit of $26~\rm mag\,arcsec^{-2}$ was imposed on the groups' mean surface brightness and the minimum projected distance to the nearest galaxy was set to be three times the radius of the smallest circle containing all members \citep{Hickson1982}. Further studies based on additional images \citep{Hickson1989} and dynamical properties of galaxies \citep{Hickson1992} have refined the original catalogue, while ensuring that member galaxies satisfy the above selection criteria. To date, the properties of these groups have been extensively investigated in several studies, with a few claiming that some HCGs may be embedded within larger, looser groups \citep[e.g.,][]{Ribeiro1998,Tovmassian2006,Mendel2011,Zheng2021}. For example, kinematical analyses led by \citet{DeCarvalho1997} and \citet{Ribeiro1998} and based on a spectroscopic survey of regions surrounding 17 HCGs, identified potential additional members of the groups, with several containing up to more than a dozen galaxies. In particular, these studies found that HCG 16, 90 and 97 respectively contain seven, nine and 14 potential galaxy members. However, as noted in \citet{DeCarvalho1997}, it is worth noting that these studies do not apply the selection criteria originally defined in \citet{Hickson1982}. Instead, they merely identified galaxies that could potentially belong to larger structures in which the compact groups are embedded. Therefore, in this study, we choose to discard the additional members listed in \citet{Ribeiro1998} and, with the exception of HCG 16 \& 31 (see \Cref{sec:census:int}), only consider those originally defined in \citet{Hickson1992}.

Furthermore, we have searched for additional members following selection criteria similar to those of \citet{Hickson1982} and \citet{Hickson1992}, but based on the $r$-band photometry derived in \Cref{sec:optical:phot} and galaxies' systemic velocities compiled from NED. We have considered a galaxy to be an additional core member if (i) its $r$-band magnitude is within a 3-magnitude range of the brightest member, (ii) its projected separation from the group's centre is less than twice the median pairwise galaxy separation, and (iii) its systemic velocity is within three times the group's velocity dispersion. Condition (i) is similar to that imposed in \citet{Hickson1982} and condition (iii) is a more conservative limit of the range of redshift distribution of 1000~\kms\, calculated in \citet{Hickson1992}. Moreover, condition (ii) was further imposed to prevent adding galaxies spatially isolated from the group. We note that these criteria are not intended to mimic those of the original studies; instead, we wish to include in the core any additional galaxy that presents a high probability of being dynamically associated to the group, rather than being a chance alignment. Of the HCGs included in this study, only two galaxies (ESO466-G044 \& ESO466-G046; see \Cref{fig:core_panels}) in the surroundings of HCG 90 satisfy these criteria. They are respectively classified as lenticular and early spiral, and located at respective systemic velocities of $\Delta v{=}183$ and $-317$~\kms\ away from the group's centre. Both galaxies were previously identified by \citet{DeCarvalho1997} as potential members (respectively denoted HCG 90-06 and 90-05 therein) of the group, along with three other galaxies located in the surrounding. However, \citet{Ribeiro1998} further suggested that HCG 90 \& 97 are likely central parts of loose groups extending beyond their boundaries \citep[see also][]{Tovmassian2006}, explaining the abundance of galaxies in the surroundings of these groups. Since no clear evidence of interaction with the rest of the core members of HCG 90 is seen in neither ESO466-G044 or ESO466-G046, we choose to not include them in the core for the rest of the present study. Instead, we classify them as outskirts, together with all other galaxies in the surroundings. We remind the reader that the purpose of this exercise is not to redefine the HCGs, but to gain insights into the distribution \hi\ content in these groups and across their evolutionary path. Therefore, we expect that a limited margin of error in the core/outskirts classification will not significantly impact the results derived from the analysis presented in this work (see \Cref{sec:add_mems}).

For each group, we define the central coordinates as the centre of mass of the group:
\begin{equation}
    {\bf P} = \frac{\sum_i m_i {\bf p}_i}{\sum_i m_i},
\end{equation}
where ${\bf p}_i$ and $m_i$ respectively represent the position and stellar mass of member galaxy $i$. The updated group centre coordinates are listed in \Cref{tab:com_coords} and their position relative to the literature values in \Cref{fig:core_panels}. The separation between the two positions is negligible for Phase 2 groups; however, for Phase 3 groups, we note a significant separation (${>}0.8'$) of the updated positions with those of the literature.

The size of the groups can be characterised in different ways: by considering the radius of the circle encompassing the centres of the galaxy members \citep[e.g.,][]{Hickson1982,DeCarvalho1997}, the median group pairwise separation \citep[e.g.,][]{Hickson1992,Ribeiro1998} or simply through the virial radius. We opt for the latter as it not only accounts for the projected separations between the individual members, but also for the distributions of their radial velocities. We first estimate the total mass of the group through the approximation of the virial mass of \citep[][equation 4]{Heisler1985}
\begin{equation}\label{eq:mvir}
    M_{\rm vir} = \frac{3\pi N}{2G}\frac{\sum_{i}^{N} (v_i - V_{\rm sys})^2}{\sum_{i<j} \frac{1}{r_{ij}}},
\end{equation}
where N is the total number of galaxy members in the group, G the gravitational constant, and $v_i$ and $V_{\rm sys}$ respectively the systemic velocities of galaxy $i$ and the group and $r_{ij}$ the pairwise projected separation between members $i$ and $j$. For simplicity, the equation assumes that all group members have equal masses, such that the group's centroid coincides with its gravitational centre. Of course, this is a simplification of the actual picture of the dark matter halo encompassing the galaxies in the groups. However, consequently to the selection criteria imposed on the HCG galaxies, the core members span at most an order of magnitude in stellar mass (see \Cref{tab:hcg_mem_props}; the groups that exhibit the largest variation in member $\log{M_\star}$ values are HCG 30 \& 97, with ranges of 1.1 and 1.0 dex, respectively). Furthermore, as noted in \citet{Heisler1985}, the approximation provides good mass estimates even if the galaxies have a range of masses. To evaluate the uncertainty on the total mass, we perform a simple error propagation on \Cref{eq:mvir} in which we neglect the errors on the pairwise separations $r_{ij}$ (the positions of the member galaxies are determined with high accuracies), which gives
\begin{subequations}\label{eq:mvir_err}
\begin{equation}
\delta_{M_\text{vir}} = M_{\text{vir}} \frac{\delta_{X_v}}{X_v}
\end{equation}
with 
\begin{equation}
X_v = \sum_i^N (v_i - V_{\rm sys})^2,
\end{equation}
\begin{equation}
\delta_{X_v} \approx 2 \sqrt{\sum_i (v_i - V_{\text{sys}})^2 \delta_{v_i}^2 + \left[\sum_i(v_i - V_{\rm sys}) \right]^2\delta_{V_\text{sys}}^2}.
\end{equation}
\end{subequations}

Here, $\delta_{v_i}$ and $\delta_{V_\text{sys}}$ are respectively the uncertainties on $v_i$ and $V_\text{sys}$. We assume a 2\% error on $v_i$ (more conservative than the typical values reported on NED) and a choose the error on $V_\text{sys}$ to be the standard error of the mean velocity, that is, $\sigma_v/\sqrt{N}$ ($\sigma_v$ here denotes the group's velocity dispersion). Next, from the total mass, we follow the virial theorem and estimate the radius as
\begin{subequations}\label{eq:rvir}
\begin{equation}
    r_{\rm vir} = \left(\frac{G\,M_{\rm vir}}{100\,H_0^2}\right)^{1/3}
\end{equation}
and its estimated uncertainty as
\begin{equation}
    \delta_{r_{\rm vir}} = \frac{r_{\rm vir}\,\delta_{M_\text{vir}}}{3\,M_{\rm vir}}.
\end{equation}
\end{subequations}
The values of $M_{\rm vir}$ and $r_{\rm vir}$, as well as their respective errors, are given in \Cref{tab:hcg_props}. It is important to note that HCGs are likely not fully relaxed and virialised structures \citep[e.g.,][]{DaRocha2008}. As a consequence, our estimate of $r_{\rm vir}$ should be interpreted as a measure of the scale of the influence of the HCG environment, rather than a measure of the actual virial radius of the HCG host halo.


\section{\hi\ Census in Hickson Compact Groups (HCGs)}\label{sec:census}

\subsection{Separating \hi\ features}\label{sec:census:sep}
The intermediate phase groups contain large \hi\ envelopes encompassing several of their members, making it difficult to trace the extent of their gas disc. To estimate the approximate \hi\ content in the discs of the groups' individual galaxy members, we follow the method of \citet{Jones2023} to attempt a feature separation. This involves creating manual 3D masks of the galaxy discs thanks to the visualisation tool {\sc SlicerAstro} \citep{Punzo2017}. The \hi\ discs of all galaxies in Phase 2 groups were successfully separated from the envelope, except for HCG 91D which was not detected. As for HCG 31, the limited angular size of the group made it impossible to distinguish the discs of its members in the $60''$-resolution datacube. Thus, we performed the separation on the highest-resolution ($15''$) cube, enabling the extraction of five core members from the envelope (HCG 31A, B, C, G \& Q).


\subsection{\hi\ masses}\label{sec:census:mass}


The \hi\ mass is calculated from the total line flux density $S\Delta v$ and luminosity distance $D$ as
\begin{equation}\label{eq:himass}
    \frac{\Mhi}{\Mo} = 2.36\e{5}\,\left(\frac{D}{\rm Mpc}\right)^2\frac{S\Delta v}{\rm Jy\,km\,s^{-1}}.
\end{equation}

The line flux is measured from the total intensity map (zeroth moment) produced by \sofia\footnote{Version 2, \url{https://gitlab.com/SoFiA-Admin/SoFiA-2}} \citep[\hi\ Source Finding Application;][]{Serra2015a,Westmeier2021} and the distance taken from \citet{Jones2023}. The line flux measurement for individual galaxies was straightforward, as the extent of their \hi\ disc is unambiguous. However, for galaxies whose \hi\ discs were connected in \hi, the manually-constructed 3-dimensional masks discussed in \Cref{sec:census:sep} were used to identify emission belonging to the discs.

There are, however, several galaxies that were not detected in the MeerKAT \hi\ data. For these, two possibilities arise: either they are devoid of atomic gas, or their \hi\ content is not sufficiently high to be detected in the MeerKAT observations. Assuming the second hypothesis, we determine the maximum \hi\ content possible for these galaxies. This content, referred to as the detection limit, is evaluated as follows: first, we build a noise map of the observed region by computing the median of the noise cube produced by \sofia\ at each spatial position. Then, we calculate the noise level at the spatial position of the galaxy by considering an area of the noise map the size of a beam and a velocity width equivalent to 20 \kms. Finally, we derive the \hi\ mass corresponding to the $3\sigma$ noise using \Cref{eq:himass}.

\begin{figure}
    \hspace{-15pt}
    \includegraphics[width=1.15\columnwidth]{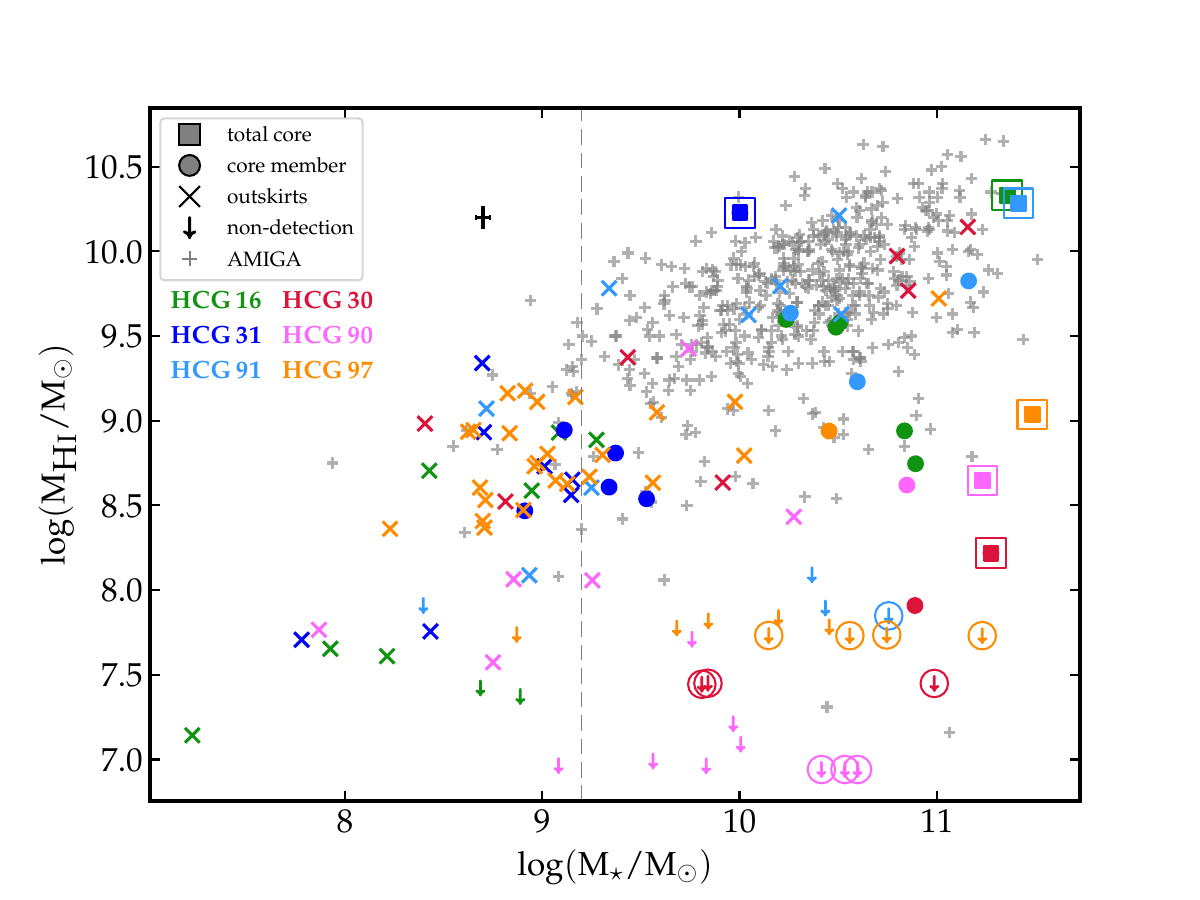}
    \vspace{-20pt}
    \caption{The \hi-stellar mass relation of the galaxies in the compact group regions (colour symbols) and of the isolated galaxies of the AMIGA sample (gray crosses). The vertical dashed line at 9.2~dex shows the lower mass limit of the AMIGA galaxies.}
    \label{fig:mhi_ms}
\end{figure}

\Cref{fig:mhi_ms} shows the \hi-stellar mass relation for galaxies in the vicinity of all six groups, compared to a sample of 518 isolated galaxies \citep{Bok2020}. These galaxies were selected from the Analysis of the interstellar Medium in Isolated GAlaxies \citep[AMIGA;][]{Verdes2005a} project, a well-defined sample of strictly isolated galaxies drawn from the Catalogue of Isolated Galaxies \citep[CIG;][]{Karachentseva1973}. The isolation criteria imposed to AMIGA galaxies are more conservative than those considered in the original CIG sample, based on two isolation parameters $\eta$ and $Q$ evaluating the surface density of neighbours out to the 5th nearest neighbour and the tidal force exerted by neighbouring galaxies, respectively. Multi-wavelength studies of the AMIGA galaxies, conducted over the past two decades, have demonstrated that the sample exhibits the lowest levels of all properties that are enhanced by interactions \citep[e.g.,][]{Lisenfeld2007,Leon2008,Lisenfeld2011,Espada2011,Jones2018,Sorgho2024}. This property establishes AMIGA as the most reliable baseline in the literature for evaluating the normalcy of \hi\ content of galaxies and assessing the influence of environmental interactions on this content.

For each of the groups, the total mass of the \hi\ emission contained in the core of the group is also measured (coloured squares in \Cref{fig:mhi_ms}). This includes both the \hi\ from the galaxies making up the core and that in the IGrM medium. Because most Phase 3 group members (warm-coloured filled circles) are undetected in \hi, their estimated total core \hi\ masses are more than an order of magnitude lower than those of Phase 2 groups (cool-coloured filled circles). However, no systematic difference is observed in the \hi\ masses of the outskirts galaxies of both phases (coloured crosses).

\begin{figure*}
    \centering
    \includegraphics[width=1\textwidth]{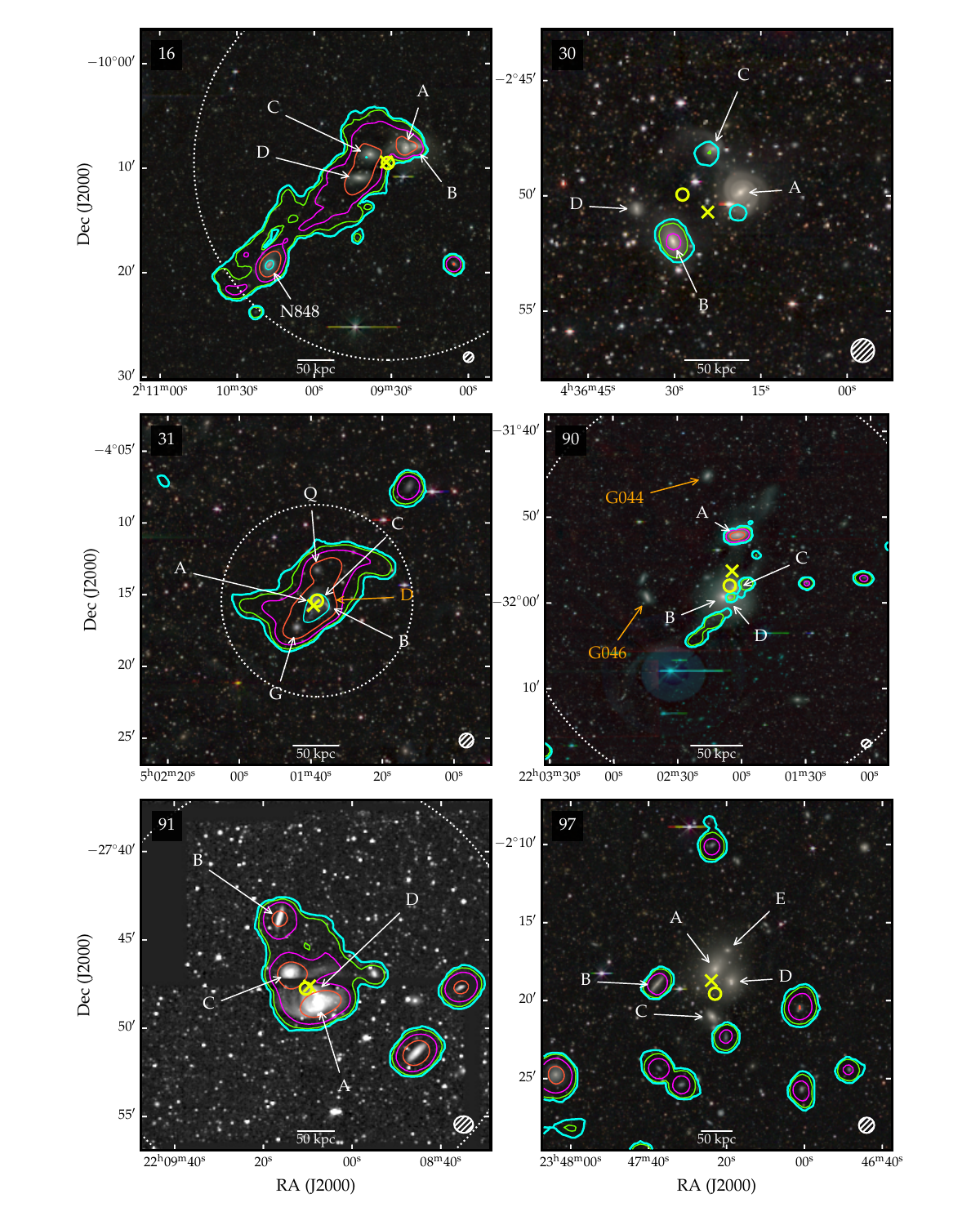}
    \vspace{-30pt}
    \caption{\hi\ contours of the central regions of the Phase 2 (left) and 3 (right) HCGs overlaid on their optical {\it grz} colour images from DECaLS (except for HCG 91 where the only DECaLS $g$-band image is shown). The contours are at $N_{\textsc{Hi},3\sigma}\times 2^{2n}$, with $N_{\textsc{Hi},3\sigma}$ given in \Cref{tab:hcg_obs} and $n=0,1,2,3$ respectively for the {\it cyan, green, magenta} and {\it orange} contours. The core galaxies are labelled with white letters, and the yellow circle and cross respectively show the NED and calculated positions of the group centres. The objects G044 and G046 in HCG 90's panel correspond respectively to ESO466-G044 and ESO466-G046 (see \Cref{sec:optical:centres} for details). The dotted circle shows the group's virial radius where it can be seen. The HCG number is given in the upper left corner of each panel and the synthesised \hi\ beam is represented by a hatched ellipse in the lower right corner.}
    \label{fig:core_panels}
\end{figure*}

\subsection{Phase 2 or intermediate-phase groups}\label{sec:census:int}
Earlier studies showed that intermediate phase groups still possess a significant amount of gas in their cores \citep{Verdes2001,Jones2023}. The left panel of \Cref{fig:core_panels} shows the \hi\ distribution in the core of the three observed HCGs, and the \hi\ masses of their members are summarised in \Cref{tab:mass_members}. Below we give a brief description of the individual groups and their surrounding galaxies. We note that the search for surrounding galaxies are limited within a maximum separation of 1000~\kms\ from the HCG systemic velocities to avoid including foreground and background galaxies.

\input{tables/tab_mems_phase2_3}

\subsubsection{HCG 16}\label{sec:census:int:h16}
The core of HCG 16 is formed by five galaxies (HCG 16A, 16B, 16C, 16D and NGC 848) embedded in an envelope spanning ${\sim}24'$ across, or a projected distance of ${\sim}342$ kpc, with a total \hi\ mass of $\rm 2.1\e{10}\,\Mo$. The fifth member, NGC 848, was not initially included in \citet{Hickson1982}'s catalogue, but was identified by \citet{DeCarvalho1997} and was later observed to be connected to the four other members in \hi\ \citep{Verdes2001}. In projection, the early spirals (Sa) 16A and 16B form the closest pair, although they are ${>}200$~\kms away from one another. Despite its high infrared luminosity and close projected proximity to 16B, early H$\alpha$ studies found that 16A presents no signs of recent interactions in its inner disc \citep[e.g.,][]{Mendes1998}. At an angular distance of ${\sim}3'$ east of the pair, lies the third member 16C – itself located ${\sim}2'$ northwest of 16D. The most distant member of the core, NGC 848, is a barred early spiral of a similar size that lies ${\gtrsim}10'$ southeast of 16D. All galaxies making up the system are thought to either be starburst-dominated or have active galactic nuclei \citep[AGNs;][]{Ribeiro1996,Turner2001}, with the pair 16A and 16B confirmed to be hosting AGNs \citep[][]{OSullivan2014,Oda2018}. Outside this collection of galaxies and within a radius of $30'$ (${\sim}428\rm\,kpc$), seven \hi-bearing galaxies were detected with \hi\ masses ranging from $\rm 1.4\e{7}\,\Mo$ to $\rm 8.5\e{8}\,\Mo$. All these detections are within 147~\kms\ of the group's systemic centre, except for the object ``PGC 4584000'' located 814~\kms\ away (${\sim}8$ times the velocity dispersion) in the foreground.

\subsubsection{HCG 31}\label{sec:census:int:h31}
Unlike in HCG 16, the objects forming the HCG 31 group are actively star-forming galaxies exhibiting signs of recent or ongoing interactions \citep[e.g.,][]{Iglesias1997,Amram2007,AlfaroCuello2015}. 
The group has a core mass of $\rm 1.7\e{10}\,\Mo$ spread over five galaxies (HCG 31A, B, C, G and Q) and several tidal features. Initially, HCG 31 was thought to host four members \citep[31A, B, C \& D;][]{Hickson1982}; however, as our understanding of its structure improved thanks to the availability of multi-wavelength data, the group was reconfigured to include new members \citep{Rubin1990} and reject 31D \citep[which happens to be a background galaxy\footnote{the approximate position of HCG 31D is given in \Cref{fig:core_panels}};][]{Hickson1992}.
Morphologically, the members of HCG 31 comprise three late spirals (A, B \& C) and two irregulars (G \& Q) spanning the velocity range 3990 -- 4136 \kms. Apart from these main galaxies, there is a complex of three emission line, star-forming regions E, F and H \citep{Rubin1990,Gallagher2010,GomezEspinoza2023} in the southern region of the group, associated with a prominent \hi\ tail \citep{Verdes2005}. The current MeerKAT observations reveal that the \hi\ structure encompassing the members of the group spans an angular size of ${\sim}8.9'$, translating to a projected distance of ${\sim}137\rm~kpc$ at the group's distance. In a $30'$-radius (${\sim}462$~kpc) of the group's centre five galaxies were detected in \hi; we also detected two additional galaxies beyond this radius and within a maximum separation of ${\sim}825$~kpc. The \hi\ masses of all seven surrounding galaxies range from $\rm 5.1\e{7}\,\Mo$ to $\rm 2.2\e{9}\,\Mo$, with a maximum radial separation of 438~\kms.

\subsubsection{HCG 91}\label{sec:census:int:h91}
HCG 91 is a quartet of late-type galaxies (91A$-$D), dominated by the barred face-on spiral, 91A, interacting with a fainter companion, 91D. Moreover, optical images of 91A show the presence of a prominent tidal tail in the galaxy pointing towards 91C \citep{Eigenthaler2015}. Early studies of the group have revealed that both these members host \ha-emitting regions, presenting highly disturbed velocity fields in their inner discs \citep{GonzalezDelgado1997,Amram2003}. Further spectroscopic studies of 91C have detected rapid variations of oxygen abundance in the arms of the galaxy, providing evidence for gas infall and interstellar medium (ISM) enrichment along 91C's arms \citep{Vogt2015,Vogt2017}. The core of the group is similar to those of the other two intermediate phase groups in terms of \hi\ content: it has a mass of $\rm 1.9\e{10}\,\Mo$ distributed over an envelope of length ${\sim}10'\rm~(267~kpc)$. The envelope extends to a cloud located west of the group, whose peak coincides with the optical object LEDA 749936. However, since no redshift information is available for the optical counterpart, we do not include it in the core. One of the four members, 91D, overlapping in projection with 91A, is not detected in \hi\ \citep{Jones2023}. Outside the core, we have detected six galaxies within a $30'$ radius (${\sim}803$~kpc) and an additional five within ${\sim}1141$~kpc. The maximum largest velocity separation with the group's centre is 381~\kms. The \hi\ masses of the detections range from $\rm 1.2\e{8}\,\Mo$ to $\rm 8.5\e{9}\,\Mo$.

\subsection{Phase 3 or late-phase groups}\label{sec:census:late}
Unlike the intermediate phase, groups in the late stage of the evolutionary sequence tend to exhibit a high \hi\ deficiency, most likely a result of their advanced evolution stage. In the right panel of \Cref{fig:core_panels} we show an overview of the \hi\ morphology in the groups and in \Cref{tab:mass_members} we list the estimated \hi\ mass of their members.

\subsubsection{HCG 30}\label{sec:census:late:h30}
The core of HCG 30 consists of four galaxies in a quadrilateral shape with a ${\sim}6'$ (about 106 kpc) diagonal: HCG 30A, 30B, 30C and 30D. Except 30C, which is identified as an intermediate-late spiral (Sbc), all galaxies of the quartet are early spirals. Previous interferometric \hi\ studies have found that HCG 30 is among the most \hi-deficient groups \citep{Verdes2001,Jones2023}, although no clear evidence for diffuse hot gas was found in its IGrM \citep{Rasmussen2008,Desjardins2014}. Unlike past observations where no member of the group was detected in \hi, we unambiguously uncover \hi\ in 30B, with a mass of $\rm 8.1\e{7}\,\Mo$. Moreover, \hi\ emission is detected for the first time at the respective spectral positions of 30A and 30C, but with a spatial offset peak with respect to their centres. 

A total of seven \hi-bearing galaxies are located around the core of the group, five of which fall inside the $30'$-radius (532~kpc) and the other two within a 866~kpc separation. Their \hi\ masses range from $\rm 3.3\e{8}\,\Mo$ to $\rm 1.4\e{10}\,\Mo$, with a maximum velocity separation of 243~\kms.

\subsubsection{HCG 90}\label{sec:census:late:h90}
HCG 90 is also a quartet of an early spiral (HCG 90A), two ellipticals (90B \& 90C) and an irregular (90D). The most ``isolated'' and northern member of the group, 90A, is a Seyfert galaxy hosting a sub-kpc cold molecular gas ring presenting characteristics of gas outflow \citep{AlonsoHerrero2023}. A multi-wavelength study of HCG 90 in the NUV and optical wavelengths enabled the discovery of five low surface brightness (LSB) dwarf galaxy candidates near 90A and 90C \citep{OrdenesBriceno2016}. Chandra X-ray observations of HCG 90 found diffuse hot gas confined to individual members of the group and forming a bridge between 90B, C and D \citep{Jeltema2008,Desjardins2013}. Similarly to HCG 30, this group does not show X-ray emission that permeate the IGrM. Consistently with \citet{Jones2023}, our MeerKAT observations detect \hi\ in one (90A) of the four members of the group: its measured \hi\ mass is $4.2\e{8}\rm\,\Mo$. Furthermore, an \hi\ stream of mass $\rm 2.1\e{8}\,\Mo$, seemingly aligned with an optical structure, is detected across the central members of the group. This feature is discussed in \citet{Ianja2025}. Among the surrounding galaxies of the group, three were detected in \hi\ inside a $30'$ radius (288~kpc) and an additional three within ${\sim}402$~kpc (all located within 435~\kms\ of the group centre).

\subsubsection{HCG 97}\label{sec:census:late:h97}
Two ellipticals (HCG 97A \& 97D), a late spiral (97B) and two early spiral (97C \& 97E) galaxies constitute the core of HCG 97, all lying within the velocity range 6003 -- 6932 \kms. Of the six groups studied here, HCG 97 is the least isolated: it has dozens of neighbours within its virial radius and has the largest velocity dispersion (${\sim}342$~\kms). The group is detected in both UV and infrared \citep{Lenkic2016}, and X-ray observations show that it has a significantly extended X-ray halo, exhibiting a morphology indicative of recent or ongoing galaxy-galaxy interactions \citep{Mulchaey2003,Rasmussen2008,Desjardins2014}. Similarly to the previous Phase 3 groups, we find that all galaxies but one are devoid of \hi\ in the group core. The neutral gas-bearing galaxy 97B, has a measured \hi\ mass of $\rm 8.7\e{8}\,\Mo$. Recently, \citet{Hu2024} conducted a multi-wavelength study in radio continuum, X-ray and CO(2-1) to demonstrate that 97B is the subject of a ram-pressure stripping scenario, forming a radio emission tail pointing towards the centre the of the group. Outside the group core, we have detected 25 galaxies in the MeerKAT observations within ${\sim}742$~kpc ($30'$) and 855~\kms\ from the centre. An additional galaxy was detected at a projected separation of ${\sim}1058$~\kms.




\section{\hi\ deficiency of galaxies in HCGs}\label{sec:hidef}



\subsection{Estimating the \hi\ deficiency}\label{sec:hidef:est}
\hi\ deficiency, as a measure of the deviation of a galaxy's \hi\ mass from its counterparts, is an important parameter for quantifying the environmental effects on the galaxies' gas content. It is defined as the difference between the galaxy's predicted and actual (or measured)
\hi\ masses:
\begin{equation}
    {\rm def_\hi} = \log(M_\hi^{\rm pred}) - \log(M_\hi^{\rm obs})
\end{equation}
In this definition, $\rm def_\hi > 0$ denotes an \hi-deficient galaxy while a gas-rich galaxy possesses a negative $\rm def_\hi$ value.

To estimate the predicted \hi\ mass, we used the almost ``nurture-free'' AMIGA sample of isolated galaxies. Thanks to their strict isolation criteria, this sample represents the best baseline for \hi\ normalcy in the absence of interactions. An $M_\hi{-}L_{\rm B}$ scaling relation was derived for the AMIGA sample in \citet{Jones2018}, enabling the estimate of the expected \hi\ mass of a galaxy within 0.25 dex. However, since we have access to the $g$- and $r$-band magnitudes of the HCG galaxies rather than their $B$-band magnitudes, we re-derived the scaling relation for the $r$-band:
\begin{equation}\label{eq:mpred}
    \log(M_\hi^{\rm pred}/\Mo) = \alpha^T\left[\log(L_r/{L_\odot}) - 10\right] + \beta^T,
\end{equation}
where $L_{r}$ is the $r$-band luminosity of the galaxy and $\alpha^T$ and $\beta^T$ are constants dependent of the RC3 \citep{DeVaucouleurs1991} morphological type $T$. Similarly to \citet{Jones2018}, we define three main morphological classes: $T{<}3$, $3{\leq}T{\leq}5$ and $T{>}5$. Of the 844 galaxies included in the \citet{Jones2018} study, 738 have DECaLS photometric measurements in the SGA  \citep[Siena Galaxy Atlas;][]{Moustakas2020} catalogue. These were retrieved through a cross-match between the \citet{Jones2018} sample and the SGA catalogue based on the spatial coordinates. We selected the $r$ magnitudes of the galaxies at the 26th isophote, which were converted into luminosity via
\begin{equation}\label{eq:mag_to_lum}
    \log{(L_{r}/L_\odot)} = 10 + 2\log{(D/{\rm Mpc})} + 0.4(M_{r,\odot} - r).
\end{equation}

The \hi\ masses measured in \citet{Jones2018} are from single-dish observations, presumably more sensitive to extended features than interferometric measurements. Of the 738 retrieved galaxies, the \hi\ masses of 205 could not be properly measured; for these, only upper limits were given \citep[see details in][]{Jones2018}. To be consistent with the analysis conducted therein and derive an equivalent \hi\ scaling relation for the $r$-band, we similarly account for the upper limits in the linear regression. However, we adopt here a Bayesian approach (instead of the maximum likelihood estimate approach used in \citealt{Jones2018}) consisting of incorporating both the detections and non-detections, separately, into a likelihood function. We used the \verb+Python+ implementation of the Markov Chain Monte Carlo {\sc PyMC} \citep[version 5.16.2;][]{AbrilPla2023}, in which we define the \Cref{eq:mpred} as a linear regression model. The distribution of the detections was modelled as a Normal distribution, whereas for the non-detections we used a Censored distribution defined as
\begin{equation}
f_{Y_{\text{c}}}(y) = \begin{cases} 
      \mathcal{N}(\mu,\sigma) & y < M_{\hi,\rm lim}, \\
      1 - F_Y(\mu_{\rm c}, M_{\hi,\rm lim}) & y = M_{\hi,\rm lim}.
   \end{cases}
\end{equation}
Here, $\mu = \alpha^T (x-10) + \beta^T$ models, following \Cref{eq:mpred}, the mean distribution of the predicted \hi\ mass for a given $x$ and $\sigma^T$ its associated intrinsic scatter. Furthermore, $F_Y(\mu_{\rm c},M_{\hi,\text{lim}})$ represents the probability that a normally-distributed variable with ``censored'' mean $\mu_{\rm c}$ is below the \hi\ mass limit $M_{\hi,\rm lim}$ in the \citet{Jones2018} data. The formulated distribution thus captures the likelihood of observing uncensored values below the \hi\ mass limit and the cumulative probability of the above values.

\begin{figure}
    \hspace{-10pt}
    \includegraphics[width=1.1\columnwidth]{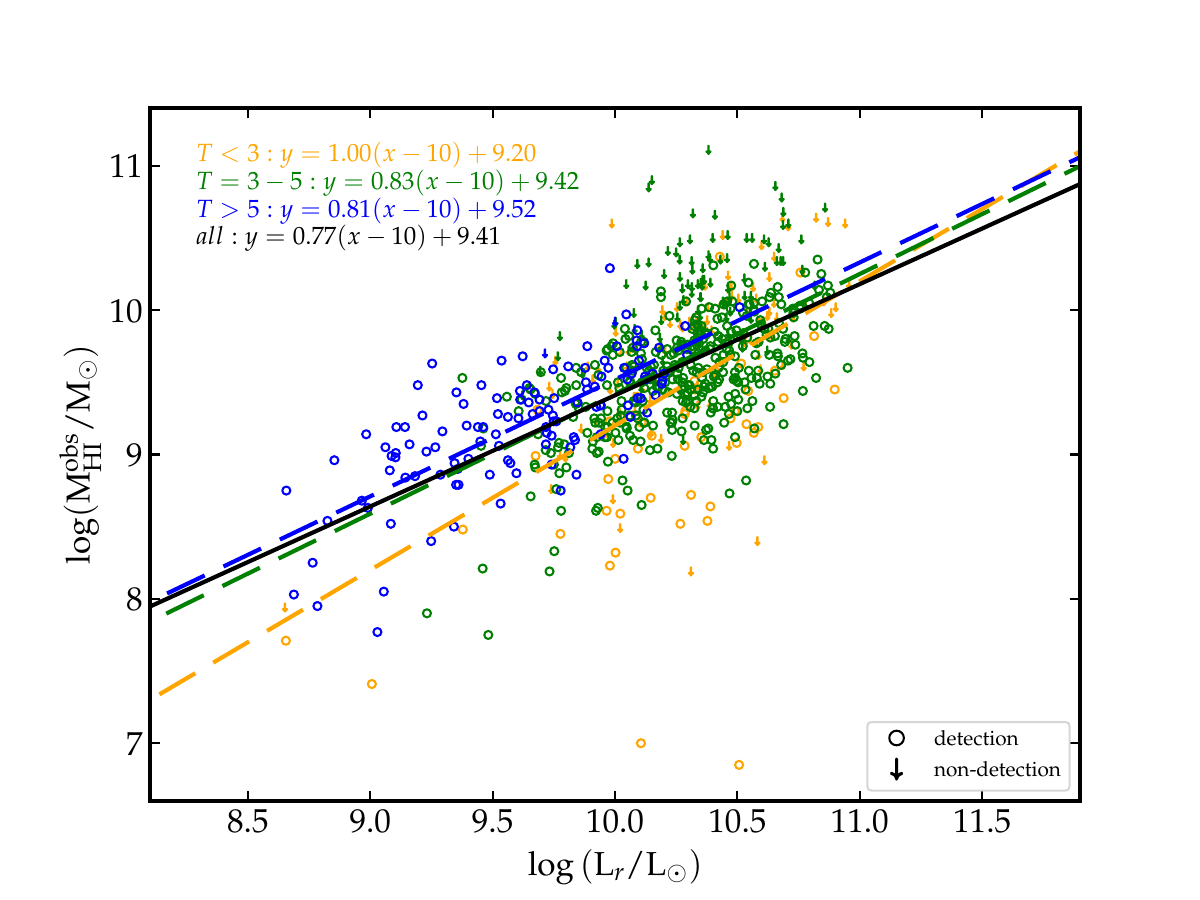}
    \caption{Predicted \hi\ mass as a function of $r$-band luminosity of the AMIGA galaxies. The open symbols represent the measured \hi\ masses while the downward arrows denote galaxies for which upper limit masses were considered. The sample is divided into morphology bins, whose fits are shown in dashed lines of corresponding colours. The fit of the overall sample is given by the solid black line.}
    \label{fig:lum_mass}
\end{figure}

For each of the free parameters $\alpha^T$ and $\beta^T$, we chose a Normal distribution of mean $\mu_\mathcal{N}=0$ and standard deviation $\sigma_\mathcal{N}=10$ as prior. The large standard deviation was chosen to probe a wide parameter space and ensure that the model is over-constrained. Additionally, the intrinsic scatter $\sigma^T$ was modelled by an Exponential distribution of coefficient 1. In short, the prior distributions are 
\begin{equation}
    \alpha^T \sim \mathcal{N}(0,10) \quad; \quad \beta^T \sim \mathcal{N}(0,10) \quad; \quad \sigma^T \sim \text{Exp}(1).
\end{equation}
Finally, 4000 Markov chains were randomly drawn to determine the posterior, from which the best-ﬁt values of the regression parameters were derived. This was done for each of the three morphology bins, and the resulting parameters are summarised in \Cref{tab:mcmc_fit}. We present the scatter plot of the $M_\hi{-}L_r$ of the AMIGA galaxies in \Cref{fig:lum_mass}, and the posterior distribution for each of the morphological bins in \Cref{fig:post_dist}.

\input{tables/mcmc_fit.tex}

\begin{figure*}
    \centering
    \includegraphics[width=1.\textwidth]{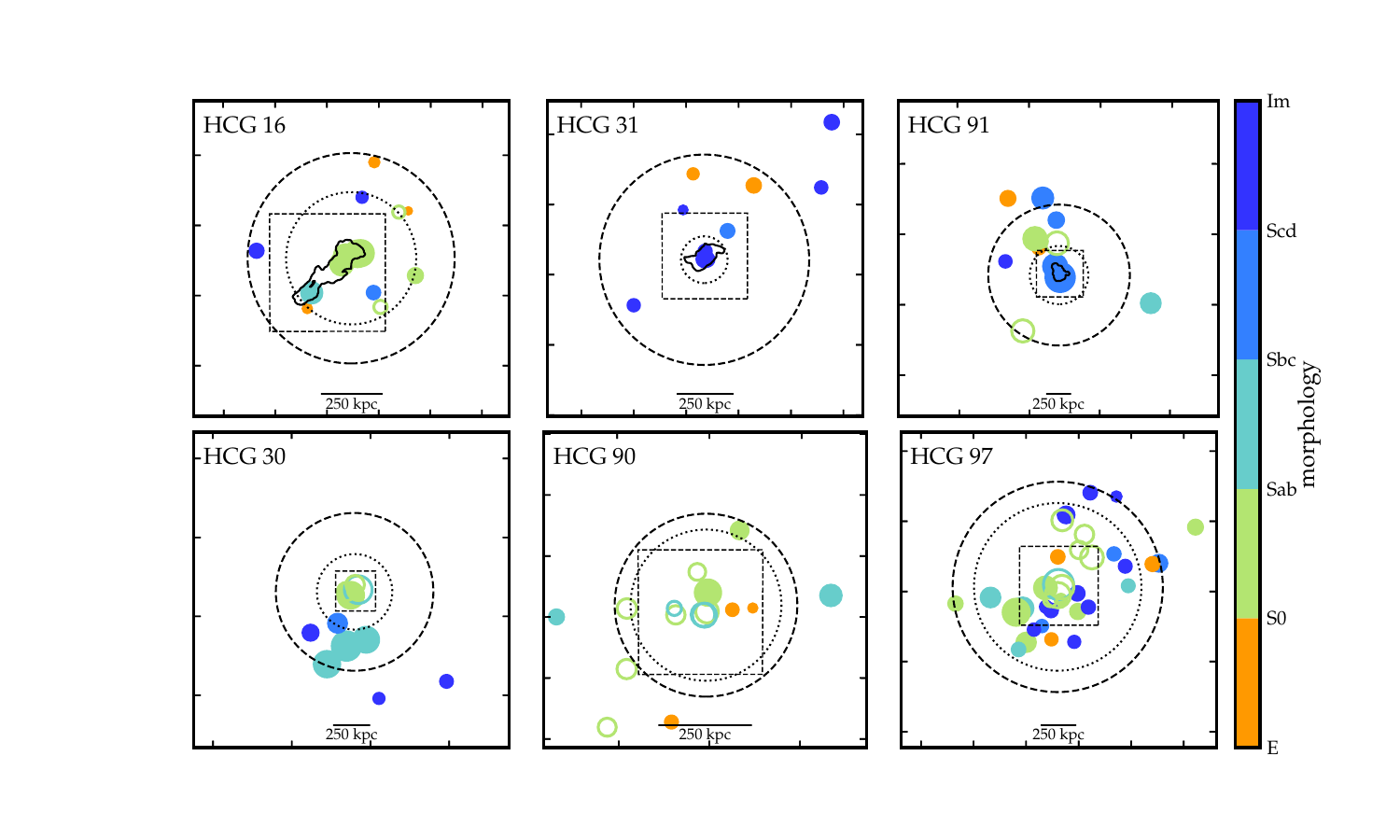}
    \vspace{-20pt}
    \caption{Morphology distribution in and around the HCGs. In the Phase 2 groups (top panels) are shown the outermost contours of \Cref{fig:core_panels}. The dashed box corresponds to the area covered in \Cref{fig:core_panels} and the dotted and dashed circles show respectively the virial radius $r_{\rm vir}$ and the $30'$ FWHM of MeerKAT. The size of the symbols is proportional to the stellar mass. Filled symbols represent sources detected in \hi\ while open circles are non-detections.}
    \label{fig:ttype_panels}
\end{figure*}

Using \Cref{eq:mpred} and the parameters in \Cref{tab:mcmc_fit}, we estimated the expected \hi\ masses of the galaxies in the HCGs based on their morphology. These morphologies were taken from the HyperLEDA\footnote{\url{http://atlas.obs-hp.fr/hyperleda}} \citep{Makarov2014} and NED databases. For galaxies that lack a morphological type in the literature, we performed a visual classification based on their DECaLS colour images. \Cref{fig:ttype_panels} shows the distribution of the morphologies in the vicinity of the compact groups, for galaxies with optical counterparts.

A striking limitation of the AMIGA sample is that it mainly includes high-mass galaxies, such that very few galaxies of $M_{\rm star}{\lesssim}10^{9.2}\,M_\odot$ are found in the sample. This is because a systemic velocity cut of $v_{\rm sys}{>}1500\rm\,km\,s^{-1}$ was imposed in the construction of the sample, biasing it against low-mass galaxies. Unfortunately, this bias translates into an under-prediction of the gas content of low-mass galaxies in the HCG galaxies. This is a direct consequence, as demonstrated in several studies \citep[e.g.,][]{Maddox2015,Hunt2020}, of the break in the \hi-to-stellar mass relation at stellar mass orders of ${\sim}10^9\,M_\odot$, where galaxies transition between gas-dominated and stellar-dominated baryonic discs. Galaxies below this mass range exhibit higher \hi-to-stellar mass fractions than more massive galaxies. To account for these systematic differences between low- and high-stellar mass galaxies, we employ a separate method to estimate the predicted \hi\ mass of our sample galaxies below $10^{9.2}\,M_\odot$. For this, we make use of the \hi-to-stellar mass relation of \citet{Bradford2015}:
\begin{equation}\label{eq:mpred_lm}
\log{(M_\hi^{\rm pred}/M_\odot)} = \begin{cases}
1.05\,\log{(M_{\star}/M_\odot)} + 0.09 & M_{\star} {<} 10^{8.6}\,M_\odot,\\
0.46\,\log{(M_{\star}/M_\odot)} + 5.18 & \text{otherwise}.
\end{cases}
\end{equation}
The relation was derived from a sample isolated galaxies selected from the NASA Sloan Atlas \citep{Blanton2011} catalogue, focusing on 368 low-mass galaxies spanning the stellar mass range $10^{7}-10^{9.5}~M_\odot$. We note that \citet{Bradford2015} use a different definition of isolation than AMIGA, solely based on its separation from its closest neighbour. Specifically, they define a low-mass galaxy ($M_\star{<}10^{9.5}\,M_\odot$) as isolated if it is located more than 1.5 Mpc from a massive host. Their analysis revealed a break in the \hi-to-stellar mass relation at stellar masses of $10^{8.6}\,M_\odot$ (just below the range found in \citealt{Maddox2015}), which the authors attribute to differences in internal processes (such as star formation efficiencies and gas temperatures) of low- and high-mass galaxies. In total, 45 of the 135 galaxies in our HCG sample have stellar masses ${<}10^{9.2}\,M_\odot$, whose predicted $\Mhi$ were hence estimated by \Cref{eq:mpred_lm}. These galaxies are essentially surrounding objects, with the exception of a core galaxy (HCG 31Q). The \hi\ deficiencies of the individual HCG galaxies, as well as their optical properties, are given in \Cref{tab:hcg_mem_props}. Likewise, those of the surrounding galaxies (including non-detections) are listed in \Cref{tab:outskirts_props}. To summarise three stellar mass bins were considered in the evaluation of the predicted \hi\ mass:
\begin{itemize}
    \item for $M_\star{<}10^{8.6}\,M_\odot$, we use the first line of \Cref{eq:mpred_lm};
    \item for $10^{8.6}\,M_\odot{<}M_\star{<}10^{9.2}\,M_\odot$, we use the second line of \Cref{eq:mpred_lm};
    \item for $M_\star{>}10^{9.2}\,M_\odot$, we use \Cref{eq:mpred} with parameters in \Cref{tab:mcmc_fit}.
\end{itemize}

Similarly to \citet{Verdes2001} and \citet{Borthakur2010}, we approximate the predicted mass of a galaxy group as the sum of the predicted individual masses. This is especially important for the intermediate phase groups (HCG 16, 31 \& 91) whose central members are embedded in large \hi\ envelopes. Their core \hi\ deficiency is thus obtained by comparing the \hi\ mass contained in the core envelopes to the total predicted mass of the core members.

\begin{figure}
    \centering
    \includegraphics[width=\columnwidth]{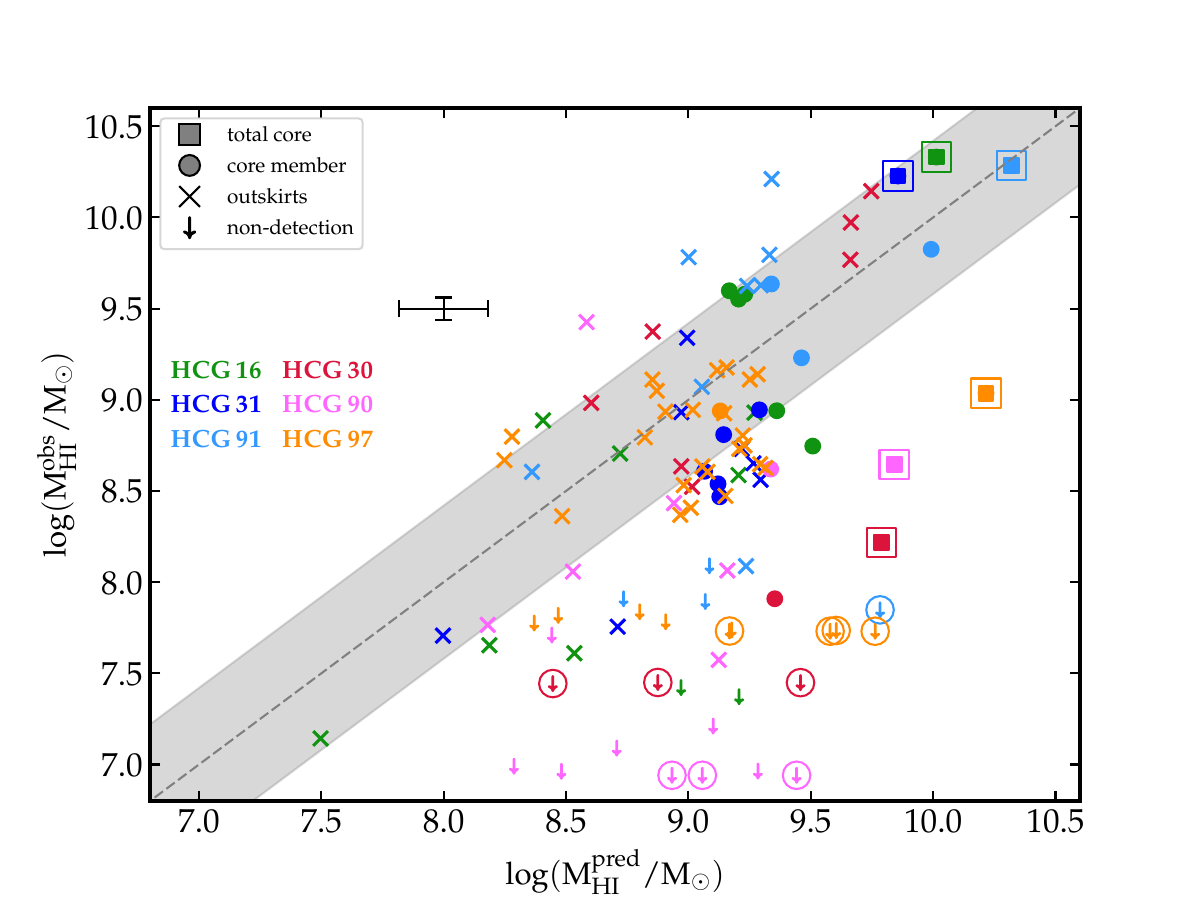}
    \caption{Predicted vs. measured \hi\ masses of galaxies in the HCGs. The dashed line and shaded region represent the one-to-one relation and intrinsinc scatter of \Cref{eq:mpred}, respectively.}
    \label{fig:mhi_obs_pred}
\end{figure}

\Cref{fig:mhi_obs_pred} shows the comparison of the measured and predicted \hi\ masses of all galaxies in a ${\sim}30'$ radius around the HCGs. Two facts are important to highlight in the figure: first, the total \hi\ contained in the cores of the late-phase groups are below the predicted values. This is in line with the expectations since most galaxies in these cores lack atomic gas (this is further shown by the individual galaxies in these cores having measured \hi\ masses below their predicted masses). On the other hand, the total measured \hi\ masses in the cores of intermediate-phase groups are within the predicted values, signifying that these cores are \hi-normal. Secondly, the outskirts galaxies around Phase 2 groups are not segregated from those around Phase 3 groups in this parameter space. We further discuss this in \Cref{sec:discussion:outskirts}.

\begin{figure}
    \centering
    \includegraphics[width=\columnwidth]{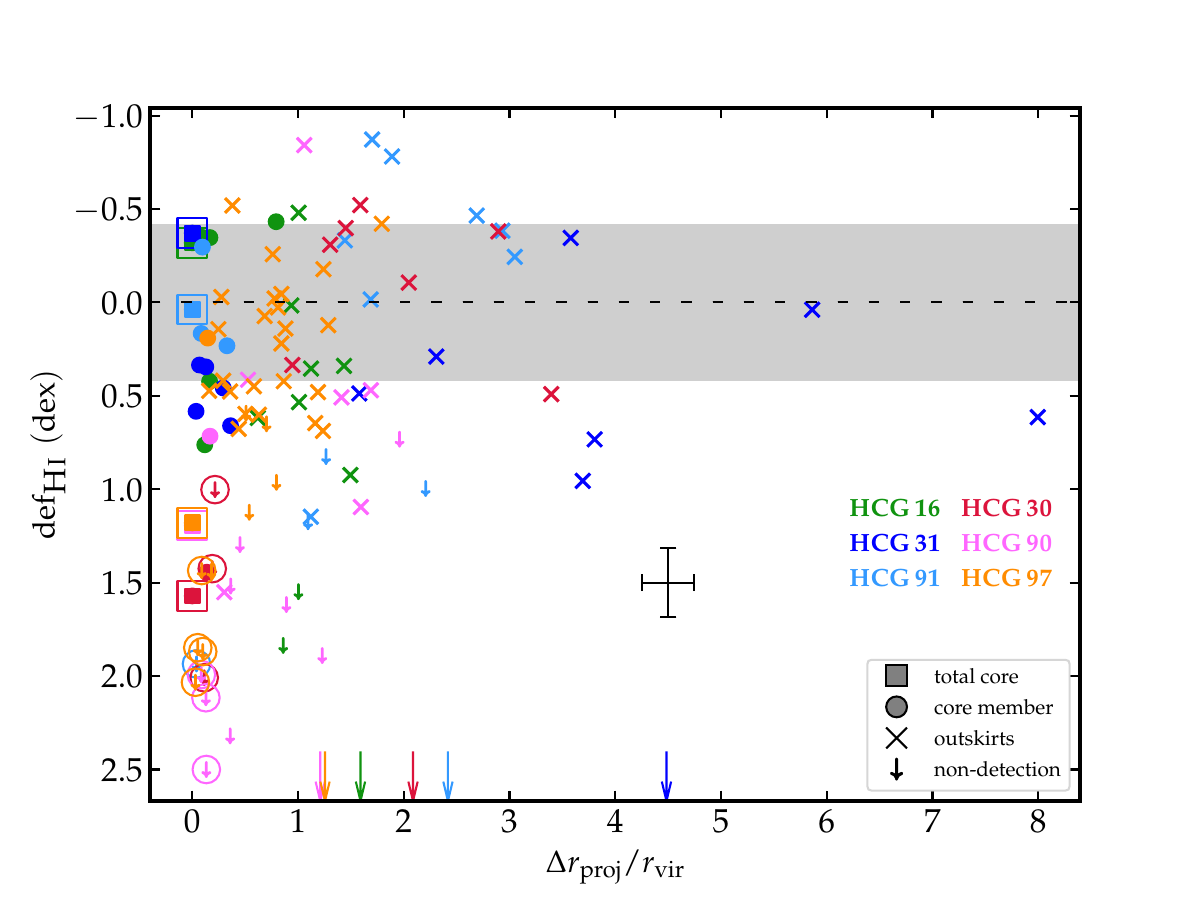}
    \caption{\hi\ deficiency vs. projected distance from group centre in units of $r_{\rm vir}$ for galaxies in and surrounding HCGs. The shaded region, corresponding to the scatter in \Cref{fig:mhi_obs_pred}, represents the zone of ``normal'' \hi\ content.}
    \label{fig:hidef_sep}
\end{figure}


\subsection{Distribution of \hi\ deficiency in HCGs}\label{sec:hidef:dist}
In their study of 38 HCGs with available \hi\ observations, \citet{Jones2023} proposed a revised evolutionary sequence of the Hickson compact groups, expanding the earlier work of \citet{Verdes2001} with an additional sub-phase in the last stage of the compact group evolution. In the additional sub-phase, a single galaxy is found to bear \hi\ in an otherwise \hi-devoid group, and where the extended features (if any) represent less than 25\% of all the detected \hi. More importantly, the authors find that while the \hi\ deficiency is similar in the early and intermediate phases of HCGs (Phase 1 \& 2), the late-phase groups exhibit deficiencies about 1.5 dex higher than the previous two. However, a clear picture of the \hi-deficiency -- or, conversely, the \hi-richness -- of the HCGs with respect to their surroundings remains to be uncovered.

Taking advantage of the MeerKAT's short baselines and large field-of-view, we aim to obtain a uniform view of the atomic gas content of the HCGs and their immediate vicinity, probing the \hi\ in galaxies out to at least one virial radius, $r_{\rm vir}$. In \Cref{fig:hidef_sep} we show the variation of the \hi\ deficiency with the projected distance (normalised by $r_{\rm vir}$) from the group centre. The first fact worth noting in the figure is the segregation between Phase 2 (cool-coloured circles) and Phase 3 core galaxies (warm-coloured circles), with the former possessing normal \hi\ content while the latter are highly deficient. This is a direct consequence of the trend observed in \Cref{fig:mhi_obs_pred}, and in line with the \hi-to-stellar mass variations of \Cref{fig:mhi_ms}. Secondly, in both phases, we observe a moderate reduction in \hi\ deficiency (indicating an increase in the atomic gas fraction) with increasing radial distance, out to one virial radius. To assess the existence of a correlation between the \defhi\ and the radius, we perform a Spearman's Rank test by calculating the correlation coefficient $\rho_{\rm def_{\hi}, r_{proj}}$ of the outskirts galaxies, i.e., of galaxies not considered as HCG members. We obtain a coefficient of ${-}0.28$ with a statistical $p{-}{\rm value}{=}1.2\e{-2}$, indicative of a weak correlation between the two variables. We further split the sample into galaxies within the virial radius and those outside, and found median \defhi\ values of $0.47{\pm}0.11\rm\,dex$ and $0.12{\pm}0.11\rm\,dex$ respectively for $r{<}r_{\rm vir}$ and $r{>}r_{\rm vir}$. We repeated the exercise for the Phase 2 and 3 groups separately, finding similar results, although with higher $p{-}{\rm values}$ (0.1 and 0.05, respectively for Phase 2 and 3) suggest a lower probability for monotonic trends. Nonetheless, a similar trend of \defhi\ values inside and outside the virial radius is observed for both phases individually.




\section{Discussion}\label{sec:discussion}
Quantifying the \hi-normalcy of a galaxy depends on the adopted definition, which may vary according to methodological preferences. For example, while several studies define \hi-normal galaxies as those having \defhi\ values within a multiple of the scatter in the \hi\ mass scaling relation (e.g., $2\sigma$ in \citealt{Solanes2002} and \citealt{Denes2014}, $1\sigma$ in \citealt{Jones2023}), others adopt a characteristic threshold value for \defhi\ (e.g., \citealt{Solanes2001} use $\rm def_\hi{>}0.3$ to quantify \hi-deficient galaxies, corresponding to two times the predicted \hi\ content).

In this study, we define a galaxy as normal if its deficiency parameter is within $\pm0.42\rm~dex$ (corresponding to the scatter of \Cref{eq:mpred} for all galaxies, see \Cref{tab:mcmc_fit}), that is, its measured mass is within the interval $0.4\,M_{\hi}^{\rm pred} < M_{\hi}^{\rm obs} < 2.6\,M_{\hi}^{\rm pred}$ ; similarly, it is deficient (rich) if \defhi\ is above (below) these limits. 

\subsection{The cores of HCGs}\label{sec:discussion:cores}
\begin{figure*}
    \centering
    \includegraphics[width=0.9\textwidth]{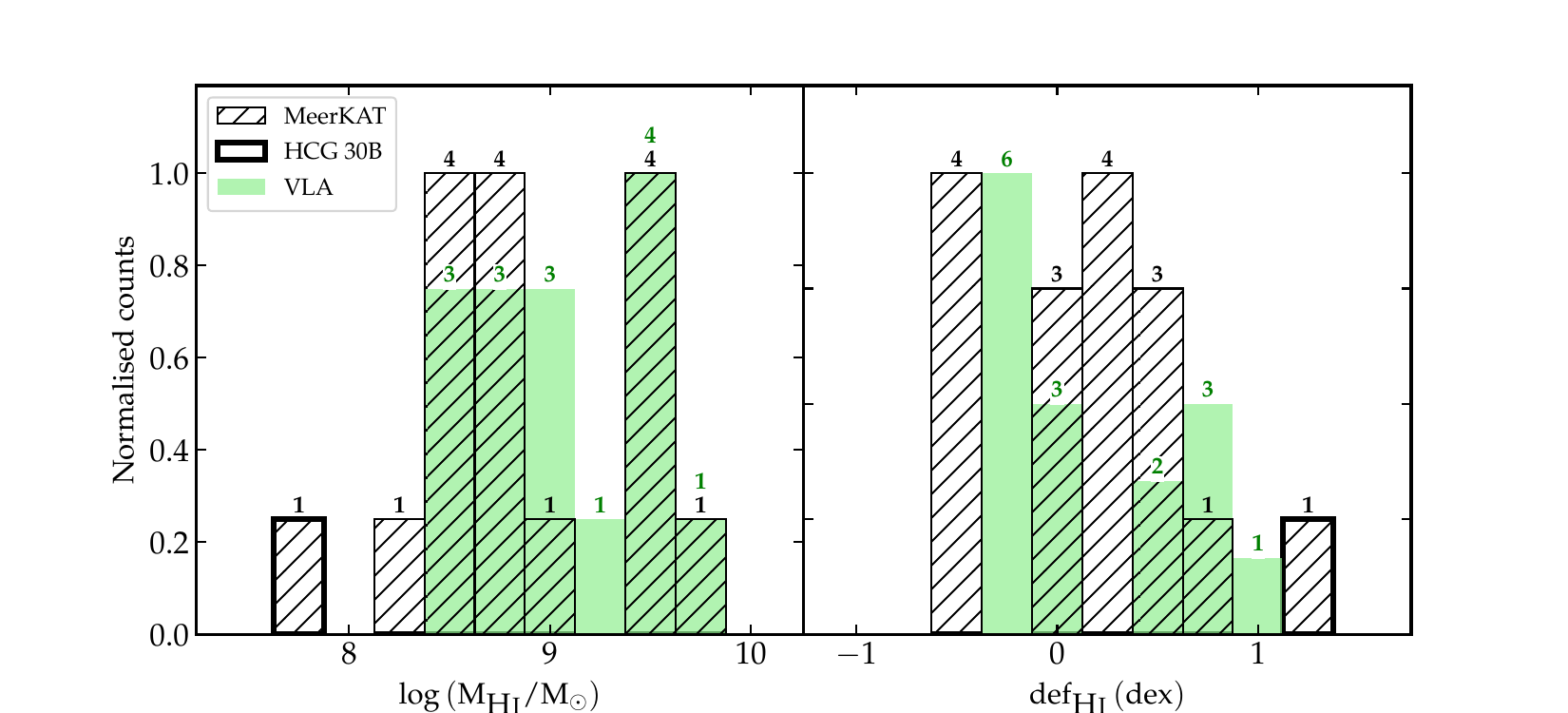}
    \caption{Comparison of the distributions of the \hi\ mass and deficiency values with the VLA measurements in \citet{Jones2023}.}
    \label{fig:hist_mkt_vla}
\end{figure*}

\citet{Jones2023} demonstrated a noticeable difference in the deficiencies of the HCG phases, with Phase 1 and 2 groups showing a similar distribution, while Phase 3 groups differ notably. Although the \defhi\ distributions peak at the same values for all group phases, that of Phase 3 groups exhibits a tail that extends to values an order of magnitude higher than those of the earlier phases. One of the goals of the present study is to revisit these results with deeper MeerKAT observations, for a subset of HCGs representative of the later two phases. In \Cref{fig:hist_mkt_vla} we show a comparison between the \defhi\ values derived in this work and those of \citet{Jones2023}. We note that these exclude the estimated limit values, since no estimate of the $\Mhi$ upper limit was given therein for non-detections. As a consequence, the majority of the objects included in the comparison are Phase 2 group galaxies; the only Phase 3 objects considered are HCG 90A and 97B. Since most of these masses were measured within manually-drawn masks (see \Cref{sec:census:sep}) which are somewhat subjective, we expect high uncertainties in the $\Mhi$ values. This is best seen in HCG 31 where, although the measured total \hi\ mass of the group is higher than that reported in \citet{Jones2023} ($1.8\e{10}\,\Mo$ versus $1.5\e{10}\,\Mo$), our combined \hi\ mass contained in the galaxies is lower ($2.5\e{9}$ versus $9.1\e{9}\,\Mo$). The remainder of the \hi\ mass is thought to belong to the IGrM in the form of bridges and tails. Nonetheless, accounting for large uncertainties, the figure suggests a general agreement in the distributions of $\Mhi$. Moreover, the scaling relation derived in this work to estimate the \defhi\ values, as well as the considered optical properties of the studied HCG galaxies, are based on recent, deep $g$- and $r$-band DECaLS photometry. However, as discussed earlier, \citet{Jones2023} made use of $B$-band magnitudes from \citet{Hickson1989}. Two main differences are worth noting: the different responses of the bands used, and the depths of the observations. The $B$ magnitudes in \citet{Hickson1989} were measured out to a surface brightness of $24.5\rm\,mag\,arsec^{-2}$, whereas the DECaLS observations produced images as deep as  ${\sim}27.9\rm\,mag\,arsec^{-2}$ \citep{Dey2019}. These differences are capable of producing discrepancies between the two measurements. Nonetheless, the mean values of the deficiencies calculated for the subsets (we remind the reader that this does not include the non-detections) are similar, with $0.30{\pm}0.03\rm\, dex$ for the MeerKAT measurements and $0.31{\pm}0.03\rm\, dex$ for the VLA results.

As discussed in \Cref{sec:census:mass}, the present MeerKAT observations allow us to set upper limits on the \hi\ masses of the non-detections. Accounting for these limits, we note a clear distinction between intermediate and late-phase groups in the global \hi\ deficiency of their cores, with Phase 2 groups presenting ``normal'' cores while Phase 3 groups are very deficient (\Cref{fig:hidef_sep}). Taken individually, the galaxies forming the cores of the groups present a similar trend in the distribution of their deficiencies, with the distribution of Phase 2 galaxies peaking in the \hi-normal region while that of late-phase galaxies peak in the \hi-deficient region (\Cref{fig:hist_hidef}). In \Cref{tab:hidef_stats} we present the statistics of the \defhi\ values; we note that a particular group makes an exception to the trend: HCG 31. Indeed, as many as four of the five of the group's members are \hi-poor (the fifth presents normal \hi\ content) although its core is globally \hi-normal. This is due to the \hi\ morphology of the group: a large quantity of IGrM \hi\ is found outside the galaxies, surpassing the combined mass of its members. A similar result was obtained in \citet{Verdes2005}. On the other hand, the only \hi\ detection in HCG 97, 97B, is the only Phase 3 group galaxy with an atomic gas content normal for its optical luminosity. The measured \hi\ mass of the galaxy is $8.7\e{8}\,\Mo$, 82\% higher than the $4.8\e{8}\,\Mo$ VLA mass reported in \citet{Jones2023}.


More generally, the median deficiency of Phase 2 group members is $\rm def_{\hi,2}{=}0.28{\pm}0.18\rm~dex$, while that of Phase 3 group galaxies (including non-detections) is $\rm def_{\hi,3}{=}1.85{\pm}0.24\rm~dex$. This corresponds to nearly a two-order-of-magnitude difference, indicating that during the transition from Phase 2 to Phase 3, the average galaxy loses almost all of its gas content. A particular example of this is seen in HCG 30. Unlike previous observations suggesting that 30B is devoid of atomic gas, the current MeerKAT campaign shows that the galaxy contains an \hi\ disc of measured mass $8.1\e{7}\rm\,\Mo$. However, compared to its optical luminosity, the galaxy remains highly gas-deficient; in fact, it is the most \hi-deficient detection of the whole sample, with $\rm def_{\hi}{=}1.4$. In other words, 30B has lost about 96\% of its atomic gas content throughout the evolution of the group. This is not surprising, given that HCG 30 constitutes the most gas-deficient of the 38 groups studied in \citet{Jones2023}. Moreover, peaks of \hi\ were also discovered at the positions of 30A and 30C, respectively. These lend support to the possibility that at least some members of the group did indeed undergo a process during which they have lost  a significant fraction of their gas content. This hypothesis is further supported by the \hi\ morphology of HCG 90. One of the four members of the group (90A) is \hi-bearing with a deficiency of 0.72 (i.e, ${\sim}80\%$ of missing \hi), likely the result of a recent infall into the group \citep{Jones2023}. However, a striking feature is observed in the group: a faint filament passing through the centre of the group, with a structure similar to that connecting members of HCG 16. The column density of the gas in the filament, with peaks at ${\sim}2.6\e{19}~\cm$, suggests that it is the remnant of a larger \hi\ complex dissipated as a result of gas removal processes.


\subsection{The outskirts of HCGs}\label{sec:discussion:outskirts}
The large field of view of MeerKAT (${\sim}1^\circ$) enables the mapping of the targeted HCGs and their surroundings, out to ${\sim}30'$. This advantage enables us to analyse the gas content of the groups within the context of their large-scale environment and to conduct a comparative study of their members relative to their surrounding galaxies.

As noted in \Cref{fig:hidef_sep} and discussed in \Cref{sec:hidef:dist}, the galaxies surrounding HCGs present a decreasing \defhi\ trend with the projected separation from the centre. Consequently, we have established that outskirts galaxies within the virial radius are ${\sim}0.3\rm~dex$ more gas-deficient than those outside.
This trend is consistent with the deficiency of those observed in galaxy clusters, where the intracluster medium affects the gas content of galaxies out to ${\sim}1{-}2\,r_{\rm vir}$ \citep[e.g.,][]{Giovanelli1985,Solanes2001,Deb2023}.

\begin{figure}
    \centering
    \includegraphics[width=0.9\columnwidth]{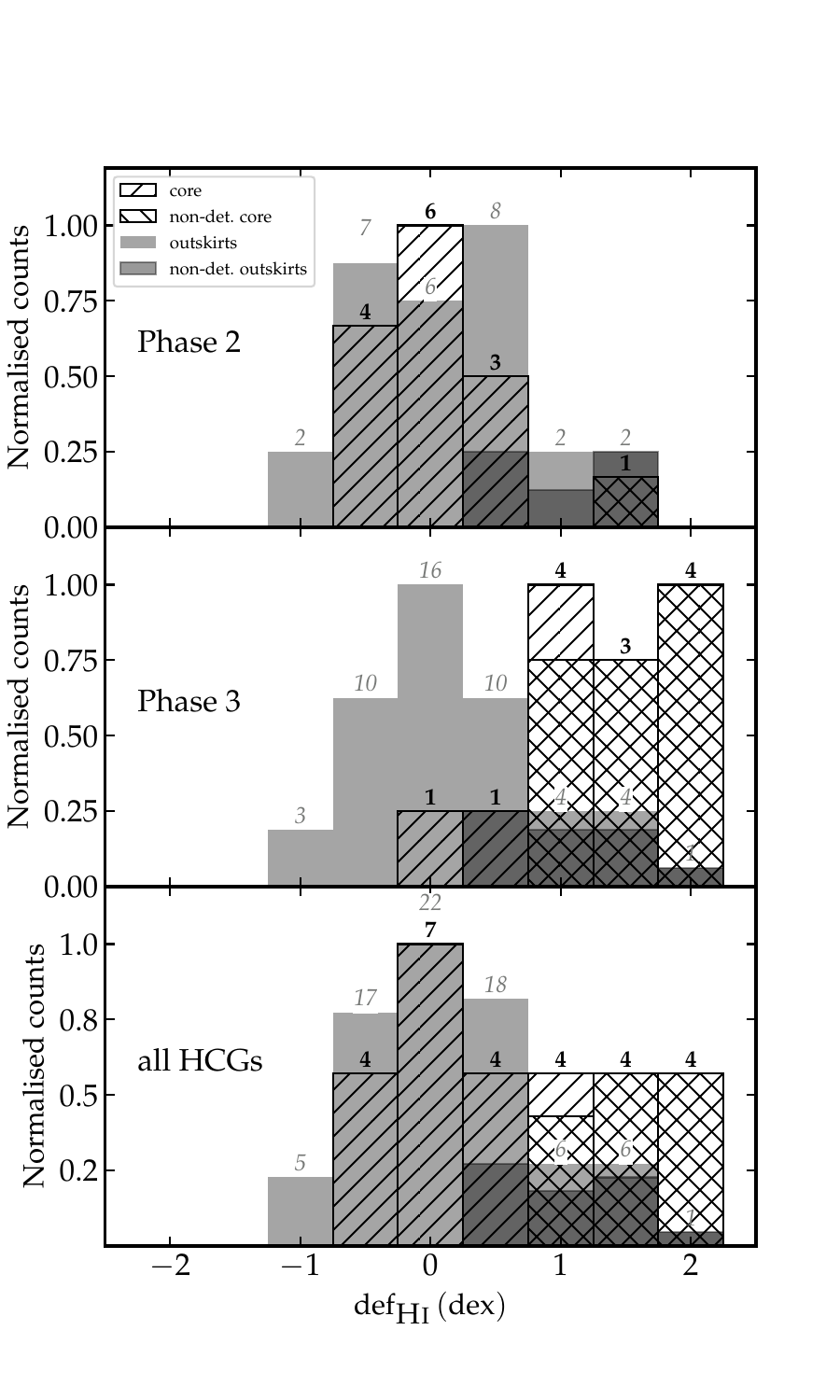}
    \vspace{-25pt}
    \caption{Distribution of the \hi\ deficiency in HCG regions for Phase 2 (top), Phase 3 (middle) and all (bottom) groups.}
    \label{fig:hist_hidef}
\end{figure}

In this section, we compare the outskirts to core galaxies in terms of gas content, independently of their separation from the group centre. We first consider all HCGs together, independently of their phase (bottom panel of \Cref{fig:hist_hidef}), and find that they do not significantly differ from their surrounding galaxies. Although the \defhi\ histograms seem to suggest that core galaxies exhibit higher deficiencies than their surroundings, their median values are consistent within the errorbars: $0.72{\pm}0.32\rm\,dex$ for core members and $0.42{\pm}0.11\rm\,dex$ for outskirts galaxies. To further verify this, we performed a Kolmogorov-Smirnov (KS) test on the cumulative distributions of \defhi\ to test the probability that the core and outskirts galaxies are drawn from the same parent population. The test provided a low degree of difference of $\rm ks{=}0.20$ and significance $p{=}0.36$, well above the 0.05 significance threshold, indicating a strong similarity between the two distributions with no significant difference. This agreement between the two distributions suggests that, globally, there are no statistical differences between HCGs and their environment in terms of \hi\ deficiency.

Next, we separate the groups by their evolutionary stage to further investigate the \defhi\ distributions (top and middle panels of \Cref{fig:hist_hidef}). We note that Phase 2 group galaxies exhibit deficiencies similar to their surrounding neighbours (the median \defhi\ values for core and outskirts galaxies are respectively $0.28{\pm}0.19\rm~dex$ and $0.29{\pm}0.23\rm~dex$). Similarly, we performed a KS test on the two distributions, obtaining a KS statistic of 0.37 and a significance level of 0.07. This is suggestive of a moderate difference between the two distributions, although the significance is marginal. On the other hand, galaxies surrounding Phase 3 groups differ significantly from the core galaxies ($\rm ks{=}0.65$ and $p{=}10^{-4}$) with deficiency values of $0.44{\pm}0.11\rm~dex$ and $1.85{\pm}0.25\rm~dex$, respectively for outskirts and core galaxies.

Although, as noted in the previous section, the galaxies in the two phases display distinct distributions, their respective surrounding neighbours show comparable \hi\ content. As previously mentioned, the six groups were selected to best represent the \hi\ morphologies of the two phases, including systematic variations from one phase to another. Despite their different \hi\ deficiencies, the large-scale environments in which they evolve present similarities in the gas distributions. Based on these facts, we argue that the evolutionary stage of an HCG is independent of the large-scale environment in which it evolves, nor does it affect the \hi\ content of galaxies outside its core. In fact, X-ray maps of HCGs in the literature \citep{Rasmussen2008,Desjardins2013} show that when available, the hot X-ray gas in the IGrM of these groups tends to be mostly confined within their cores, rarely extending beyond the farthest group member. For example, of the six groups, only HCG 97 presents an important hot gas envelope \citep[see][]{Rasmussen2008}. The second group with an important quantity of X-ray emission is HCG 90, whose hot gas envelope surrounds members 90B, C an D. However, as noted in \Cref{sec:census:late:h90}, the hot gas is considered unlikely to spread through the group's IGrM \citep{Desjardins2013}. We argue that the notable difference of Phase 3 HCGs with their surrounding is likely a consequence of the later morphological evolution of their members with respect to the rest of their environment \citep[see, e.g.][]{Coziol1998}.

\begin{figure*}
    \centering
    \includegraphics[width=1.\textwidth]{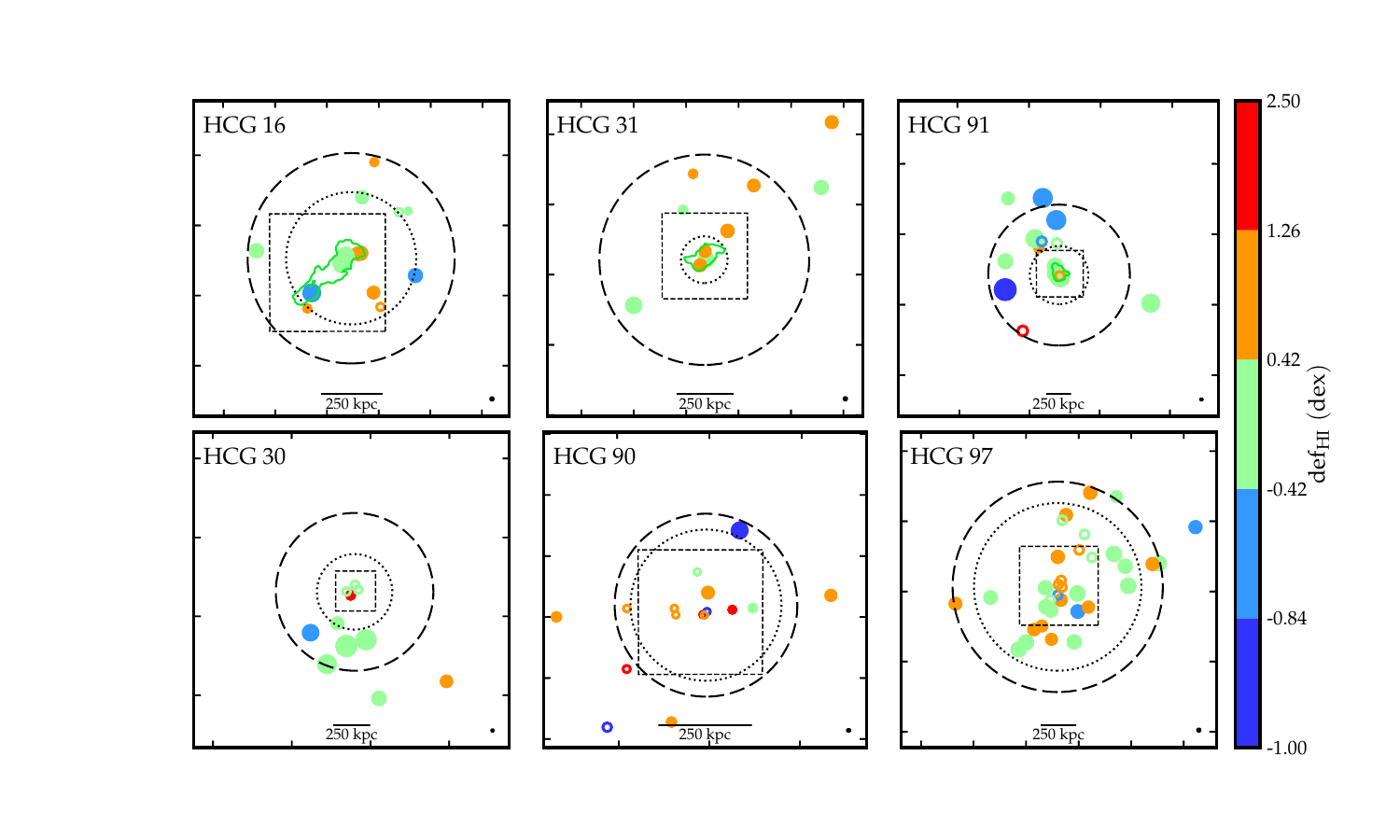}
    \vspace{-40pt}
    \caption{The spatial distribution of the deficiency parameter \defhi\ in the regions of the HCGs. Circles, symbols and contour lines are the same as in \Cref{fig:ttype_panels}. The size of the symbols is proportional to the \hi\ mass. A zoomed version of this figure, highlighting the central regions, in shown in \Cref{fig:hidef_panels_zoom}.}
    \label{fig:hidef_panels}
\end{figure*}

We further show in \Cref{fig:hidef_panels} the distribution of the galaxies in the plane of the sky for each HCG, colour-coded by the \hi\ deficiency. The \defhi\ distributions do not favour a particular phase: for example, the galaxies in the regions of HCG 30 and 91 present similar distributions (see \Cref{tab:hidef_stats}) despite the differences in the groups phases; this is also the case for HCG 31 and 97.
\setlength{\tabcolsep}{2pt}
\input{tables/hidef_stats}


\section{Summary}\label{sec:summary}
We have presented \hi\ deficiencies of galaxies in and around six HCGs, selected to represent the intermediate and late phases of their evolutionary sequence. Through a MeerKAT campaign, we aimed to update the deficiency trends observed in Phase 2 and Phase 3 groups from previous VLA observations \citep{Jones2023} and study the \hi\ content of HCGs against their immediate environment. Similarly to previous findings, we observed significant differences in the \hi\ content of the Phase 2 and 3 group galaxies. On average, Phase 3 group members are $\rm{\sim}1.5~dex$ more deficient than those of Phase 2. However, given the modest size of the sample and the large \defhi\ scatter among groups of a same phase, we note that this is not a general prescription for characterising the \hi\ content of HCG members.

Mapping the \hi\ in galaxies around the HCGs in a ${\sim}30'$ radius, we have investigated the correlation between the groups and their surrounding galaxies in terms of \defhi\ distribution. Globally, we have found that galaxies in the cores present the same distributions as their surroundings; however, when separated by their evolutionary stage, Phase 3 HCG galaxies are over an order of magnitude more deficient than their outskirts. This finding supports prior studies indicating that galaxies within evolved HCGs have undergone faster evolution compared to their surrounding regions.

\section*{Data availability and reproducibility}
All \hi\ cubes, masks and moment maps used in this work are available in \citet{Ianja2025}. Additionally, the DECaLS images of the groups, the derived photometry tables of individual galaxies as well as the separated \hi\ discs of the HCG members are available in open source in a Zenodo repository\footnote{\url{https://doi.org/10.5281/zenodo.14965528}}. The repository also includes all tables and figures featured in the paper. Furthermore, in order to make the paper as reproducible as possible, we linked the Zenodo repository to a GitHub repository containing all \verb+Python+ scripts and \verb+Jupyter+ notebooks developed as part of the analysis presented in this paper. These repositories are publicly accessible.

\begin{acknowledgements}
The MeerKAT telescope is operated by the South African Radio Astronomy Observatory (SARAO), which is a facility of the National Research Foundation, an agency of the Department of Science and Innovation.
This work used the Spanish Prototype of an SRC \citep[SPSRC,][]{Garrido2021} service and support funded by the Ministerio de Ciencia, Innovaci\'on y Universidades (MICIU), by the Junta de Andaluc\'ia, by the European Regional Development Funds (ERDF) and by the European Union NextGenerationEU/PRTR. 
Authors AS, LVM, RI, KMH, RGB, AdO, JP, SSE, JG, BN JM, TW and EP acknowledge financial support from the Severo Ochoa grant CEX2021-001131-S funded by MICIU/AEI/10.13039/501100011033. Part of the work of AS, LVM, RI, JG, SSE and TW was funded by the grant PID2021-123930OB-C21 funded by MICIU/AEI/10.13039/501100011033, by ERDF/EU.
TW acknowledges financial support from grant TED2021-130231B-I00 funded by MICIU/AEI/10.13039/501100011033 and by the European Union NextGenerationEU/PRTR.
AdO and JP acknowledge further financial support from the MICIU through project PID2022-140871NB-C21 by ERDF/EU.
R.G.B. acknowledges further financial support from grant PID2022-141755NB-I00.
JM acknowledges financial support from  grant PID2023-147883NB-C21, funded by MICIU/AEI/10.13039/501100011033.
JMS acknowledges financial support from the Spanish state agency MICIU/AEI/10.13039/501100011033 and by ERDF/EU funds through research grant PID2022-140871NB-C22 and the additional support of MICIU/AEI/10.13039/501100011033 through the Centre of Excellence Mar\'ia de Maeztu's award for the Institut de Ci\`encies del Cosmos at the Universitat de Barcelona under contract CEX2019–000918–M.
EA and AB acknowledge support from the Centre National d'Etudes Spatiales (CNES), France.
MEC acknowledges the support of an Australian Research Council Future Fellowship (Project No. FT170100273) funded by the Australian Government.
J.R. acknowledges financial support from the Spanish Ministry of Science and Innovation through the project PID2022-138896NB-C55.
OMS's research is supported by the South African Research Chairs Initiative of the Department of Science and Technology and National Research Foundation (grant No. 81737).
\end{acknowledgements}

\bibliographystyle{aa}
\bibliography{references}

\begin{appendix}
\begin{onecolumn}
\section{Tables}
For readability purposes, the names of some member galaxies of \Cref{tab:outskirts_props} were abbreviated following the following scheme: W=``WISEA J, B=``APMUKS(BJ) B'', G=``GALEXASC J'', S=``2dFGRS S'', X=``2MASX J'', J=``2MASS J''. For example, W020919.28-095201.8 denotes the source WISEA J020919.28-095201.8 and G050203.46-040132.9 the source GALEXASC J050203.46-040132.9.

\input{tables/centres_coords.tex}
\input{tables/gals_props_core}
\input{tables/gals_props_outskirts}

\section{Impacts of classifying ESO 466-G044/46 as HCG 90 core members}\label{sec:add_mems}
The systematic search for HCG members in \Cref{sec:optical:centres} has identified two galaxies, ESO 466-G044 and ESO 466-G046, as potential members of HCG 90. However, given that the galaxies present no apparent sign of interaction with the group, we have opted to not include them in the late-phase group. In this section, we present the alternate scenario of accounting these galaxies in the core. Specifically, we verify its impact on the main results of the analysis, contained in \Cref{fig:hidef_app}.

A comparison of the figure with \Cref{fig:hist_hidef} shows no notable differences. Specifically, the deficiency distribution of Phase 3 groups is still distinct from that of the surrounding galaxies (ks=0.69 and $p{=}10^{-5}$), with respective median values of $1.52\pm0.22$~dex and $0.42\pm0.12$~dex. Moreover, the core galaxies of the total sample (combined Phase 2 and 3 groups) present a median deficiency parameter of $0.76\pm0.35$~dex while the surrounding galaxies exhibit $0.41\pm0.12$~dex.

\FloatBarrier
\begin{figure}[h!]
    \centering
    \includegraphics[angle=90, width=0.85\columnwidth]{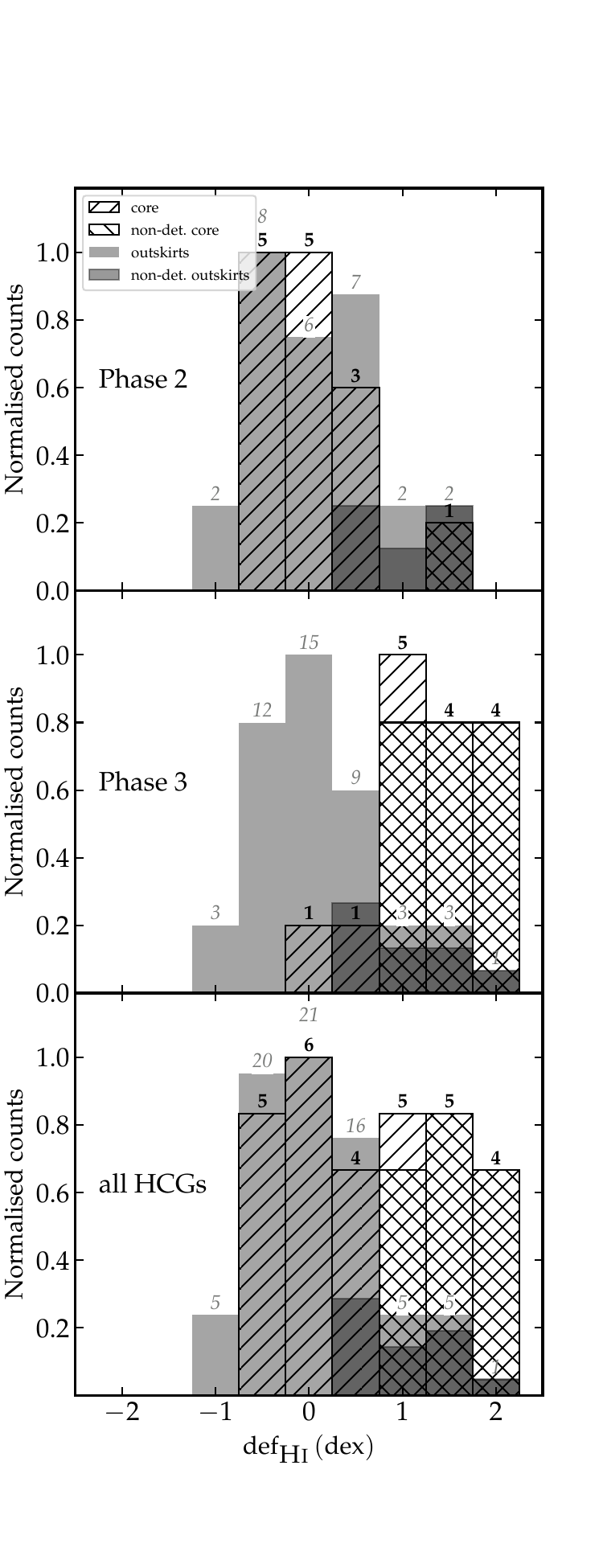}
    \caption{Same as \Cref{fig:hist_hidef} but with galaxies ESO 466-G044 and ESO 466-G046 considered as core members of HCG 90.}\label{fig:hidef_app}
\end{figure}

\section{Posterior distribution and zoom-in on the cores}
\begin{figure}[h!]
    \centering
    \begin{minipage}{0.35\textwidth}
        \includegraphics[width=\textwidth]{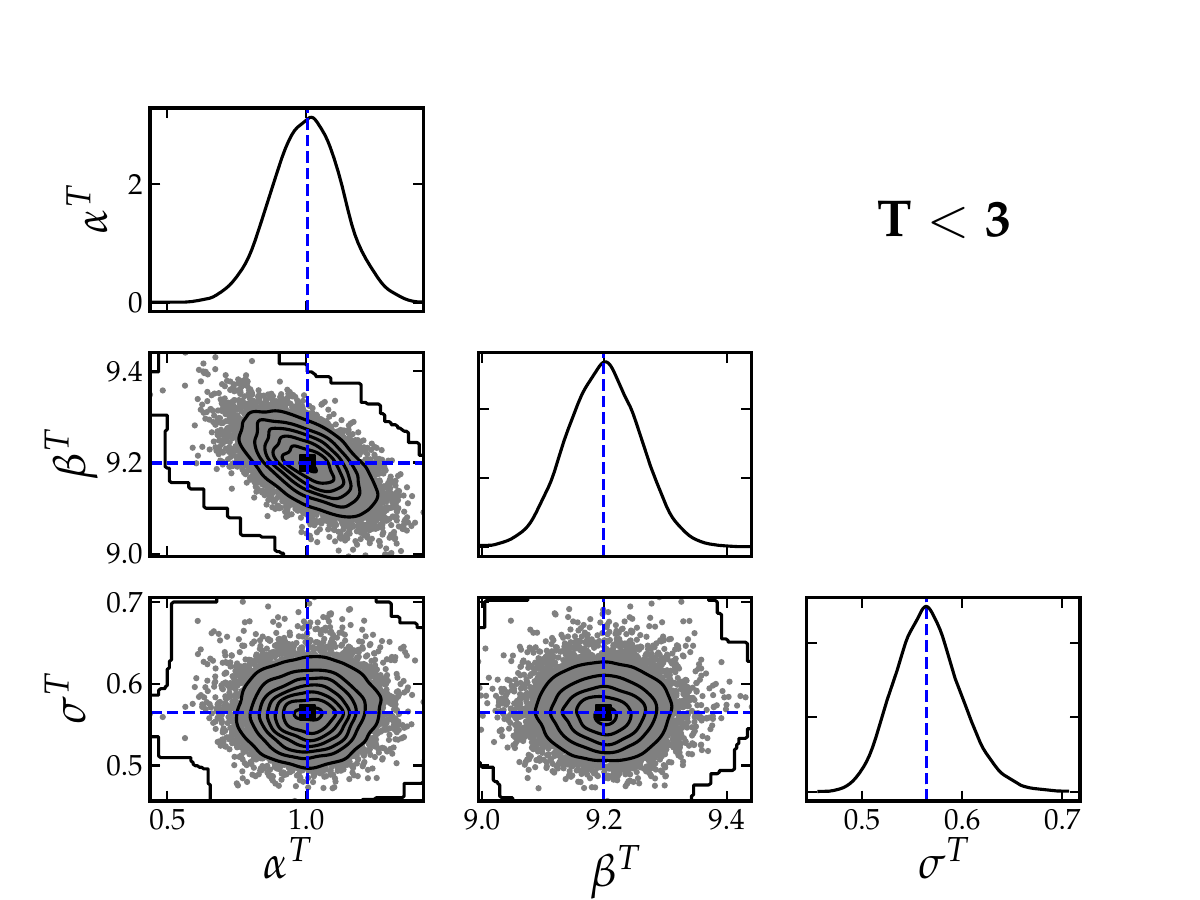}
    \end{minipage}
    \begin{minipage}{0.35\textwidth}
        \includegraphics[width=\textwidth]{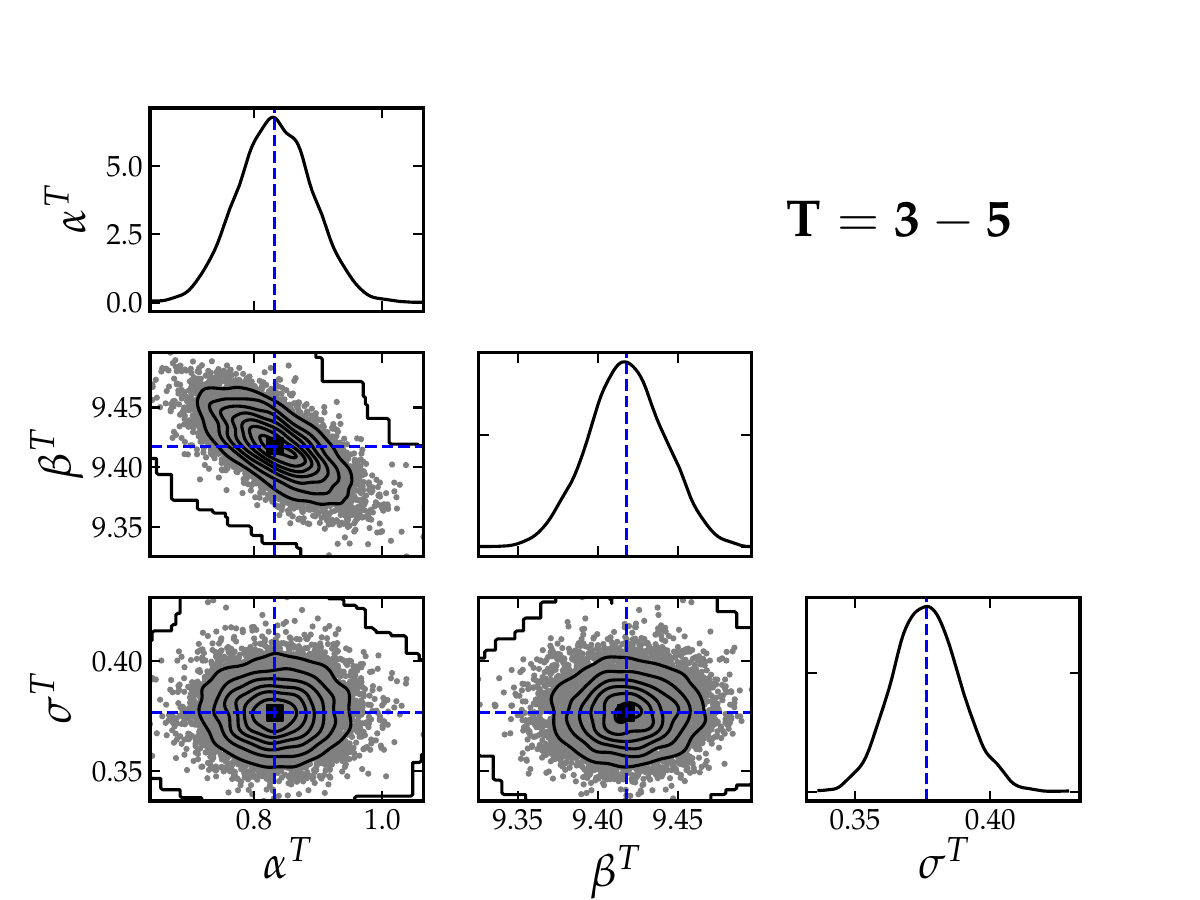}
    \end{minipage}
    \begin{minipage}{0.35\textwidth}
        \includegraphics[width=\textwidth]{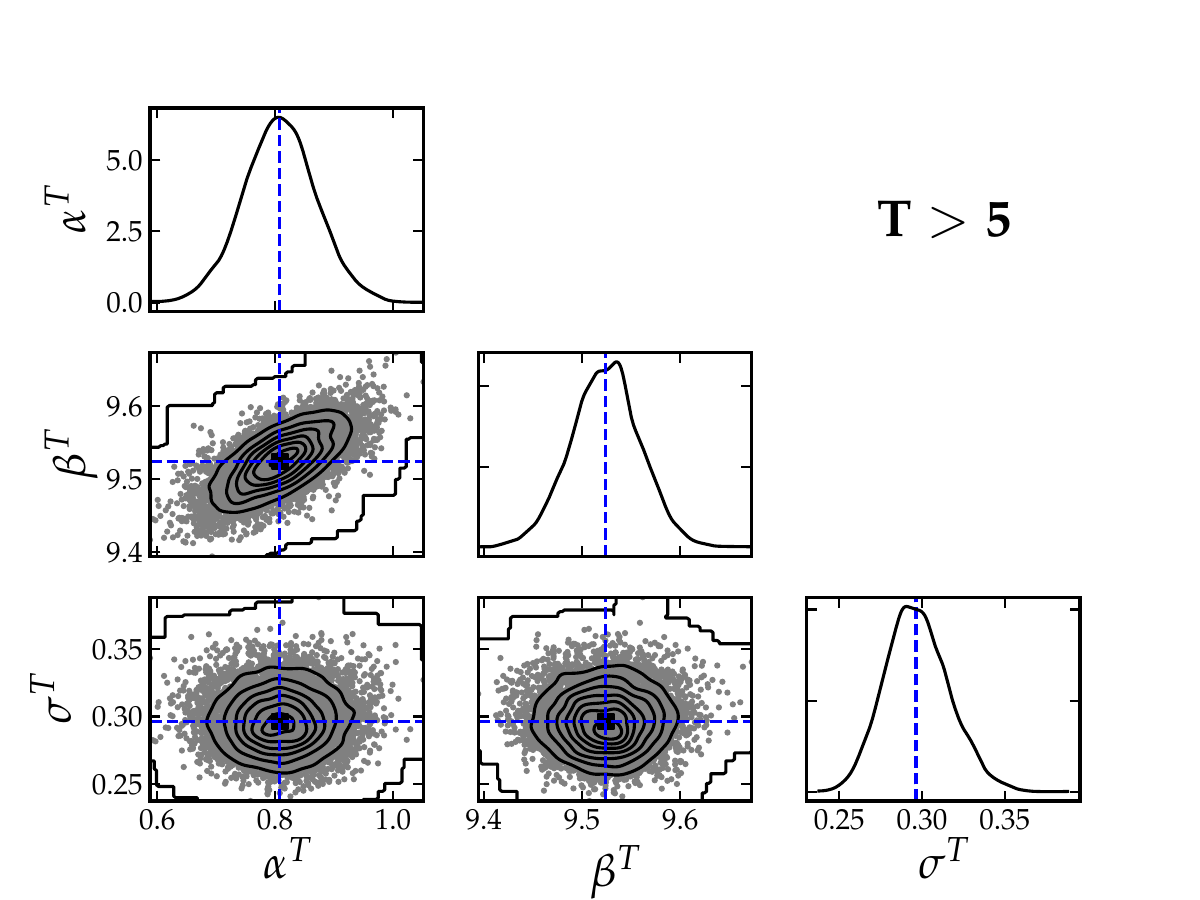}
    \end{minipage}
    \begin{minipage}{0.35\textwidth}
        \includegraphics[width=\textwidth]{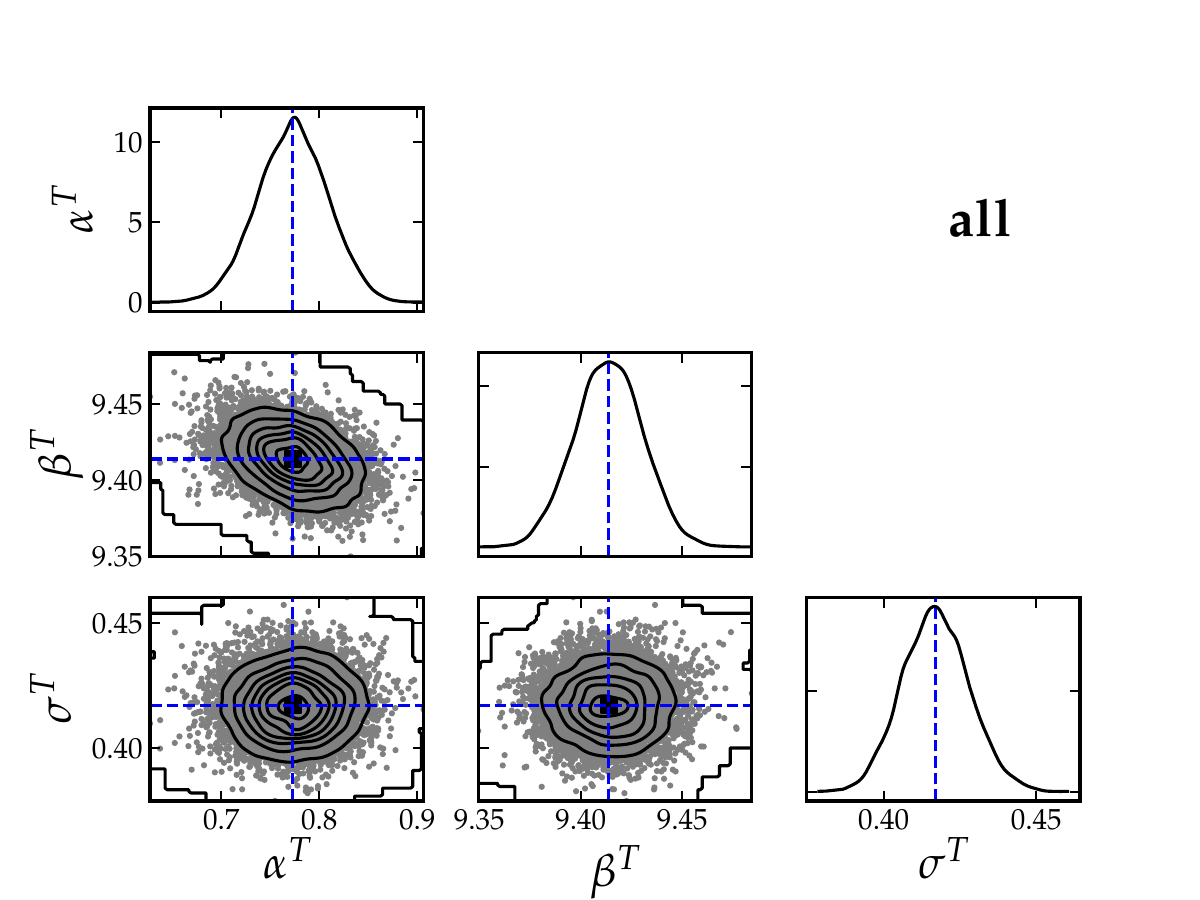}
    \end{minipage}
    \caption{The posterior distributions of the regression parameters for the AMIGA galaxies and for each of the morphological bins.}\label{fig:post_dist}
\end{figure}

\begin{figure*}
\centering
\includegraphics[width=0.7\textwidth]{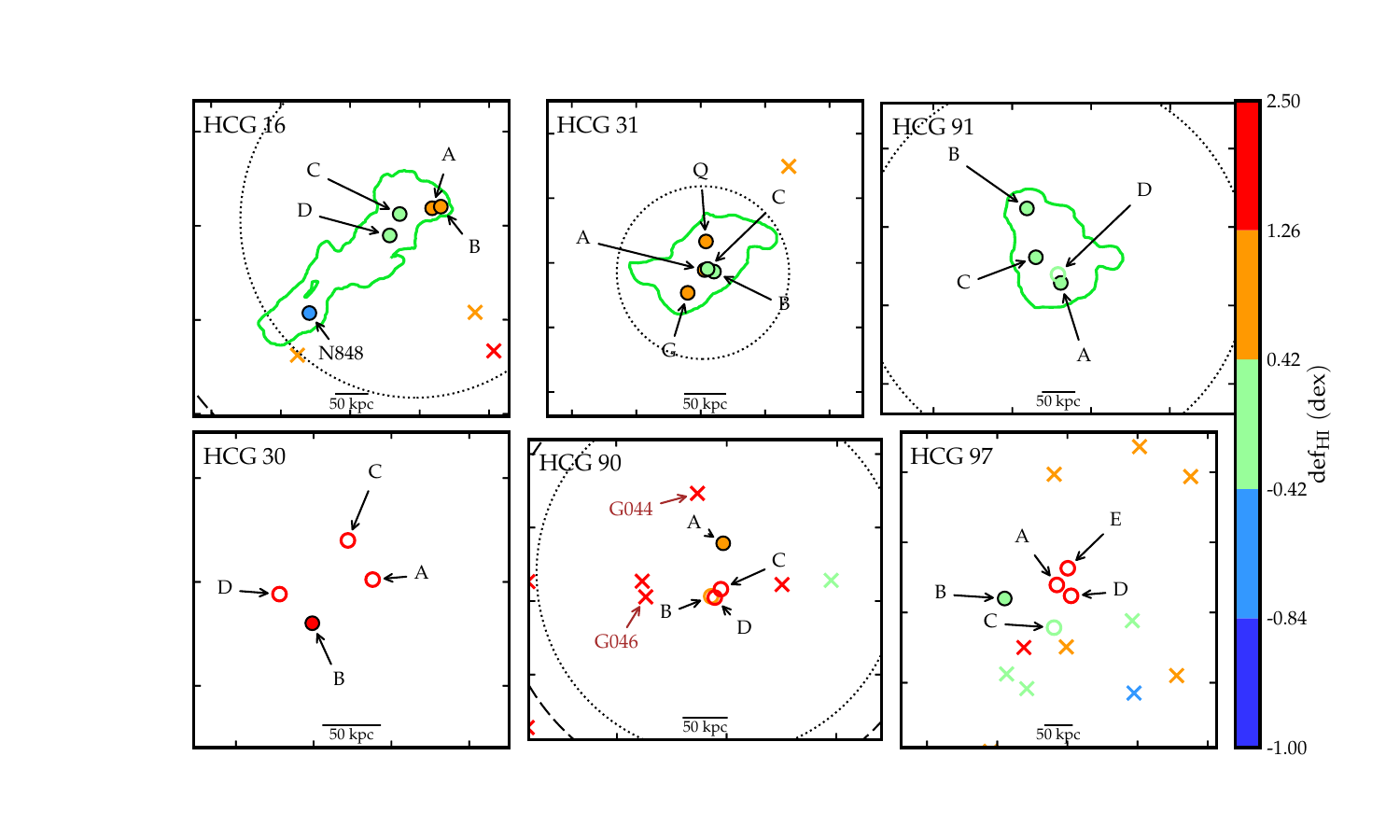}
    \vspace{-20pt}
    \caption{Central regions of \Cref{fig:hidef_panels}. The core galaxies are labelled with letters and outskirts galaxies are marked with crosses.}
    \label{fig:hidef_panels_zoom}
\end{figure*}
\end{onecolumn}
\end{appendix}
\end{document}

%% file: tables/hcg_props.tex
\begin{table*}
    {\footnotesize
    \centering
    \caption{Properties of the HCGs.}
    \label{tab:hcg_props}
    \begin{tabular}{c c c c c c c c c c}
    \hline \hline 
\multirow{2}{*}{Phase} & \multirow{2}{*}{HCG} & RA & Dec & $V_{\rm sys}$ & $\sigma_{v}$ & Dist. & $M_{\rm vir}$ & $r_{\rm vir}$ & \multirow{2}{*}{$\rm N$}\\ 
 & & \multicolumn{2}{c}{(J2000)} & ($\rm km\,s^{-1}$) & ($\rm km\,s^{-1}$) & (Mpc) & ($10^{12}\,M_\odot$) & (kpc) & \\ 
(1) & (2) & \multicolumn{2}{c}{(3)} & (4) & (5) & (6) & (7) & (8) & (9) \\ 
\hline \rule{0pt}{10pt} 
\multirow{3}{*}{2} & 16 & 02:09:31.97 & -10:09:25.67 & 3977 & 98.9 & 49 & $2.2\pm1.8$ & $268.8\pm73.7$ & 5\\ 
 & 31 & 05:01:39.27 & -04:15:45.63 & 4068 & 50.1 & 53 & $0.1\pm0.1$ & $103.1\pm28.7$ & 5\\ 
 & 91 & 22:09:09.59 & -27:47:36.70 & 7195 & 190.4 & 92 & $4.2\pm1.9$ & $332.1\pm49.3$ & 4\\ 
\hline \rule{0pt}{10pt} 
\multirow{3}{*}{3} & 30 & 04:36:24.21 & -02:50:42.39 & 4645 & 105.0 & 61 & $1.9\pm0.5$ & $255.0\pm21.1$ & 4\\ 
 & 90 & 22:02:04.47 & -31:56:17.88 & 2635 & 109.9 & 33 & $1.5\pm0.6$ & $237.9\pm29.6$ & 4\\ 
 & 97 & 23:47:23.91 & -02:18:43.08 & 6579 & 342.2 & 85 & $23.5\pm1.9$ & $591.1\pm16.2$ & 5\\ 
\hline
\end{tabular}
\tablefoot{
Columns list: (1) the HCG phase; (2) the HCG group ID; (3) the J2000 coordinates as computed from this work (see \Cref{sec:optical:centres}); (4) the heliocentric systemic velocity from \citet{Jones2023};     (5) the group's velocity dispersion; (6) the group's distance from \citet{Jones2023}; (7) The virial mass of the group; (8) the virial radius of the group based on its velocity dispersion; (9) the number of group members considered. See text for details on derived properties.}}
\end{table*}

%% file: tables/tab_mems_phase2_3.tex
\begin{table}
{\footnotesize 
\begin{center}
\caption{\hi\ masses of individual HCG core members.}\label{tab:mass_members}
\begin{tabular}{l l c c} 
\hline 
\hline 
\multirow{2}{*}{HCG} & \multirow{2}{*}{member} & $\rm\log{M_{\textsc{Hi}, gal}}$ & $\rm\log{M_{\hi,ext}}$ \\ 
 & & $\rm (M_\odot)$ & $\rm (M_\odot)$ \\ 
\hline \rule{0pt}{10pt}
 \multirow{6}{*}{16} & 16A & 8.7 & --\\ 
 & 16B & 8.9 & -- \\ 
 & 16C & 9.6 & -- \\ 
 & 16D & 9.6 & -- \\ 
 & N0848 & 9.6 & -- \\ 
 & {\bf total} & {\bf 10.1} & {\bf 9.9}\\ 
\rule{0pt}{10pt}
 \multirow{6}{*}{31} & 31A & 8.5 & --\\ 
 & 31B & 8.9 & -- \\ 
 & 31C & 8.8 & -- \\ 
 & 31G & 8.6 & -- \\ 
 & 31Q & 8.5 & -- \\ 
 & {\bf total} & {\bf 9.4} & {\bf 10.2}\\ 
\rule{0pt}{10pt}
 \multirow{5}{*}{91} & 91A & 9.8 & --\\ 
 & 91B & 9.2 & -- \\ 
 & 91C & 9.6 & -- \\ 
 & LEDA 749936$^*$ & 8.2 & -- \\ 
 & {\bf total} & {\bf 10.1} & {\bf 9.8}\\ 
\hline
\rule{0pt}{10pt}
 \multirow{5}{*}{30} & 30A & ${<}7.4$ & --\\ 
 & 30B & ${}7.9$ & -- \\ 
 & 30C & ${<}7.5$ & -- \\ 
 & 30D & ${<}7.4$ & -- \\ 
 & {\bf total} & ${\bf {\leq}8.2}$ & {\bf --}\\ 
\rule{0pt}{10pt}
 \multirow{5}{*}{90} & 90A & ${}8.6$ & --\\ 
 & 90B & ${<}6.9$ & -- \\ 
 & 90C & ${<}6.9$ & -- \\ 
 & 90D & ${<}6.9$ & -- \\ 
 & {\bf total} & ${\bf {\leq}8.6}$ & {\bf 8.2}\\ 
\rule{0pt}{10pt}
 \multirow{6}{*}{97} & 97A & ${<}7.7$ & --\\ 
 & 97B & ${}8.9$ & -- \\ 
 & 97C & ${<}7.7$ & -- \\ 
 & 97D & ${<}7.7$ & -- \\ 
 & 97E & ${<}7.7$ & -- \\ 
 & {\bf total} & ${\bf {\leq}9.0}$ & {\bf --}\\ 
\hline
\end{tabular}
\tablefoot{
The top and bottom halves respectively include the Phase 2 and 3 groups. The last two columns respectively list the \hi\ masses inside and outside the discs of the member galaxies.\\ 
$^*$LEDA 749936 (WISEA J220854.95-274701.6) coincides spatially with the peak of an \hi\ cloud in the core of HCG 91, but lacks an optical readshift; we include it in this table but do not count it as a core member.
}
\end{center}} 
\end{table}

%% file: tables/mcmc_fit.tex
\begin{table}
\centering
\footnotesize
\caption{Linear regression fit results of \Cref{eq:mpred}}\label{tab:mcmc_fit}
\addtolength{\tabcolsep}{-2pt}
\begin{tabular}{c c c c c c} 
\hline 
\hline 
\rule{0pt}{10pt}
Type & Det. & Non-det. & $\alpha^T$ & $\beta^T$ & $\sigma^T$ \\ 
\hline \rule{0pt}{10pt}
${<}3$ & 47 & 99 & $1.00\pm0.12$ & $9.20\pm0.06$ & $0.57\pm0.03$ \\ 
$3{-}5$ & 363 & 104 & $0.83\pm0.06$ & $9.42\pm0.02$ & $0.38\pm0.01$ \\ 
${>}5$ & 123 & 2 & $0.81\pm0.06$ & $9.52\pm0.04$ & $0.30\pm0.02$ \\ 
all & 533 & 205 & $0.77\pm0.04$ & $9.41\pm0.02$ & $0.42\pm0.01$ \\ 
\hline
\end{tabular}
\end{table}

%% file: tables/hidef_stats.tex
\begin{table}
{\footnotesize 
\begin{center}
\caption{Distibutions of \hi\ deficiency in galaxies in the cores and outskirts of the HCGs.}\label{tab:hidef_stats}
\begin{tabular}{c c c c c c c c} 
\hline 
\hline 
\multirow{2}{*}{Phase} & \multirow{2}{*}{HCG} & \multicolumn{3}{c}{core} & \multicolumn{3}{c}{outskirts} \\ 
\cmidrule(lr){3-5}  \cmidrule(l){6-8} 
& & \% rich & \% normal & \% poor & \% rich & \% normal & \% poor \\ 
\hline \rule{0pt}{10pt} 
\multirow{3}{*}{2} & 16 & 60 & 0 & 40 & 11 & 11 & 77\\ 
 & 31 & 0 & 0 & 100 & 14 & 14 & 71\\ 
 & 91 & $25^{}$ & $50^{}$ & $25^{a}$ & 45 & 18 & 36\\ 
\hline\rule{0pt}{10pt}
 \multirow{3}{*}{3} & 30 & $0^{}$ & $0^{}$ & $100^{b}$ & 57 & 14 & 28\\ 
 & 90 & $0^{}$ & $0^{}$ & $100^{b}$ & 8 & 0 & 91\\ 
 & 97 & $0^{}$ & $20^{}$ & $80^{a}$ & 10 & 34 & 55\\ 
\hline
\end{tabular}\\ 
\tablefoot{
$^{a}$No \hi\ detection. 
$^{b}$Only 1/4 were detected in \hi.} 
\end{center}} 
\end{table}

%% file: tables/centres_coords.tex
\begin{table*}[h!]
\centering
\caption{NED and calculated centre of mass (CoM) coordinates of the HCGs and their differences $\Delta\rm pos$.}\label{tab:com_coords}
\begin{tabular}{c c c c c c c} 
\hline 
\hline 
\multirow{2}{*}{Phase} & \multirow{2}{*}{HCG} & \multicolumn{2}{c}{NED (J2000)} & \multicolumn{2}{c}{CoM (J2000)} & $\Delta\rm pos$ \\ 
 & & RA & Dec & RA & Dec & arcmin \\ 
\hline \rule{0pt}{10pt} 
\multirow{3}{*}{2} & 16 & 02:09:31.32 & -10:09:30.64 & 02:09:31.97 & -10:09:25.67 & 0.18\\ 
 & 31 & 05:01:38.30 & -04:15:25.24 & 05:01:39.27 & -04:15:45.63 & 0.42\\ 
 & 91 & 22:09:10.43 & -27:47:45.42 & 22:09:09.59 & -27:47:36.70 & 0.24\\ 
\hline \rule{0pt}{10pt} 
\multirow{3}{*}{3} & 30 & 04:36:28.61 & -02:49:56.60 & 04:36:24.21 & -02:50:42.39 & 1.34\\ 
 & 90 & 22:02:05.62 & -31:58:00.44 & 22:02:04.47 & -31:56:17.88 & 1.73\\ 
 & 97 & 23:47:22.94 & -02:19:33.56 & 23:47:23.91 & -02:18:43.08 & 0.88\\ 
\hline
\end{tabular}
\end{table*}

%% file: tables/gals_props_core.tex
\begin{table*}[h!]
{\footnotesize
\begin{center}
\caption{Optical and \hi\ properties of the HCG members.}
\label{tab:hcg_mem_props}
\begin{tabular}{clccccccccc}
\hline \hline
\multirow{2}{*}{HCG} & \multirow{2}{*}{member} & $V_{\rm sys}$ & \multirow{2}{*}{Morph.} & $d$ & $g$ & $r$ & $\rm log{M_{\star}}$ & $\rm log{M_{\hi}}$ & $\rm def_{\hi}$ & \multirow{2}{*}{$\rm M_{\hi, limit}$} \\ 
  &  & ($\rm km\,s^{-1}$) &  & (kpc) & (mag) & (mag) & ($M_\odot$) & ($M_\odot$) & (dex) &  \\ 
 \hline\rule{0pt}{10pt} 
16 & 16A & 4075.5 & Sa & 31.6 & 13.1 & 12.3 & 10.9 & 8.7 & 0.8 & 0 \\ 
 16 & 16B & 3867.0 & Sa & 44.1 & 13.5 & 12.7 & 10.8 & 8.9 & 0.4 & 0 \\ 
 16 & 16C & 3849.0 & S0a & 25.1 & 13.7 & 13.0 & 10.5 & 9.6 & -0.3 & 0 \\ 
 16 & 16D & 3874.0 & S0a & 44.8 & 13.8 & 13.1 & 10.5 & 9.6 & -0.3 & 0 \\ 
 16 & N0848 & 3989.0 & SBab & 213.1 & 13.7 & 13.2 & 10.2 & 9.6 & -0.4 & 0 \\ 
 30 & 30A & 4697.0 & SBa & 28.7 & 13.7 & 12.9 & 11.0 & 7.4 & 2.0 & 1 \\ 
 30 & 30B & 4625.0 & Sa & 35.1 & 14.0 & 13.2 & 10.9 & 7.9 & 1.4 & 0 \\ 
 30 & 30C & 4508.0 & SBbc & 48.2 & 15.8 & 15.2 & 9.8 & 7.5 & 1.4 & 1 \\ 
 30 & 30D & 4794.0 & S0 & 54.9 & 16.2 & 15.5 & 9.8 & 7.4 & 1.0 & 1 \\ 
 31 & 31A & 4074.0 & Sdm & 3.7 & 14.9 & 14.6 & 9.5 & 8.5 & 0.6 & 0 \\ 
 31 & 31B & 4136.0 & Sm & 13.2 & 15.4 & 15.2 & 9.1 & 8.9 & 0.3 & 0 \\ 
 31 & 31C & 4037.5 & Im & 7.0 & 14.7 & 14.5 & 9.4 & 8.8 & 0.3 & 0 \\ 
 31 & 31G & 3990.5 & cI & 30.2 & 15.0 & 14.8 & 9.3 & 8.6 & 0.5 & 0 \\ 
 31 & 31Q & 4037.0 & Im & 37.4 & 16.9 & 16.5 & 8.9 & 8.5 & 0.7 & 0 \\ 
 90 & 90A & 2603.0 & Sa & 39.9 & 12.7 & 11.9 & 10.8 & 8.6 & 0.7 & 0 \\ 
 90 & 90B & 2523.0 & E0 & 30.8 & 13.4 & 12.6 & 10.5 & 6.9 & 2.1 & 1 \\ 
 90 & 90C & 2512.0 & E0 & 20.5 & 13.7 & 12.9 & 10.4 & 6.9 & 2.0 & 1 \\ 
 90 & 90D & 2775.0 & Im & 31.6 & 13.4 & 12.6 & 10.6 & 6.9 & 2.5 & 1 \\ 
 91 & 91A & 6832.0 & SBc & 28.0 & 13.3 & 12.7 & 11.2 & 9.8 & 0.2 & 0 \\ 
 91 & 91B & 7267.0 & Sc & 108.9 & 15.0 & 14.3 & 10.6 & 9.2 & 0.2 & 0 \\ 
 91 & 91C & 7285.0 & Sc & 31.9 & 15.2 & 14.7 & 10.3 & 9.6 & -0.3 & 0 \\ 
 91 & 91D & 7195.0 & SB0 & 13.1 & 14.4 & 13.7 & 10.8 & 7.8 & 1.9 & 1 \\ 
 97 & 97A & 6932.0 & E1 & 17.8 & 13.7 & 12.9 & 11.2 & 7.7 & 2.0 & 1 \\ 
 97 & 97B & 6820.0 & Sa & 86.5 & 15.2 & 14.5 & 10.5 & 8.9 & 0.2 & 0 \\ 
 97 & 97C & 6002.5 & SBd & 58.2 & 14.9 & 14.1 & 10.7 & 7.7 & 1.9 & 1 \\ 
 97 & 97D & 6327.5 & SBa & 30.8 & 13.8 & 13.3 & 10.6 & 7.7 & 1.8 & 1 \\ 
 97 & 97E & 6664.5 & SB0 & 52.5 & 16.2 & 15.5 & 10.1 & 7.7 & 1.4 & 1 \\ 
 \hline
\end{tabular}
\end{center}}
\end{table*}

%% file: tables/gals_props_outskirts.tex
\setlength{\tabcolsep}{3pt}
{\footnotesize
\begin{longtable}{clccccccccc}
\caption{Optical and \hi\ properties of galaxies surrounding HCGs.} \label{tab:outskirts_props} \\ 
\hline
\hline
\multirow{2}{*}{HCG} & \multirow{2}{*}{member} & $V_{\rm sys}$ & \multirow{2}{*}{Morph.} & $d$ & $g$ & $r$ & $\rm log{M_{\star}}$ & $\rm log{M_{\hi}}$ & $\rm def_{\hi}$ & \multirow{2}{*}{$\rm M_{\hi, limit}$} \\ 
  &  & ($\rm km\,s^{-1}$) &  & (kpc) & (mag) & (mag) & ($M_\odot$) & ($M_\odot$) & (dex) &  \\ 
 \hline\\[-1.5ex]\endfirsthead
\hline
\multirow{2}{*}{HCG} & \multirow{2}{*}{member} & $V_{\rm sys}$ & \multirow{2}{*}{Morph.} & $d$ & $g$ & $r$ & $\rm log{M_{\star}}$ & $\rm log{M_{\hi}}$ & $\rm def_{\hi}$ & \multirow{2}{*}{$\rm M_{\hi, limit}$} \\ 
  &  & ($\rm km\,s^{-1}$) &  & (kpc) & (mag) & (mag) & ($M_\odot$) & ($M_\odot$) & (dex) &  \\ 
 \hline
\\[-1.5ex]
\endhead
16 & KUG 0208-103 & 3846.0 & Sd & 385.8 & 16.0 & 15.6 & 9.1 & 8.9 & 0.3 & 0 \\ 
 16 & KUG 0206-099A & 3830.0 & S0 & 401.8 & 17.3 & 17.1 & 8.2 & 7.6 & 0.9 & 0 \\ 
 16 & W020919.28-095201.8 & 3847.0 & Sm & 251.9 & 17.6 & 17.3 & 8.4 & 8.7 & 0.0 & 0 \\ 
 16 & KUG 0205-104 & 3869.0 & S0a & 270.5 & 15.5 & 15.1 & 9.3 & 8.9 & -0.5 & 0 \\ 
 16 & KUG 0206-105 & 3972.0 & Sc & 166.6 & 16.2 & 15.9 & 8.9 & 8.6 & 0.6 & 0 \\ 
 16 & W020825.73-095553.3 & 3906.0 & S0 & 302.0 & 20.2 & 19.9 & 7.2 & 7.1 & 0.4 & 0 \\ 
 16 & W021022.68-102345.2 & 4112.0 & S0 & 271.1 & 19.3 & 18.8 & 7.9 & 7.7 & 0.5 & 0 \\ 
 16 & W020836.71-095615.7 & 4026.0 & Im & 269.9 & 18.0 & 17.4 & 8.7 & 7.4 & 1.5 & 1 \\ 
 16 & PGC 4584000 & 3163.0 & S0 & 231.1 & 18.3 & 17.5 & 8.9 & 7.4 & 1.8 & 1 \\ 
 30 & NGC 1618 & 4888.0 & SBb & 332.5 & 13.4 & 12.8 & 10.8 & 10.0 & -0.3 & 0 \\ 
 30 & NGC 1622 & 4848.0 & SBab & 370.0 & 13.0 & 12.2 & 11.2 & 10.1 & -0.4 & 0 \\ 
 30 & NCG 1625 & 4760.5 & SBb & 522.4 & 13.5 & 12.8 & 10.9 & 9.8 & -0.1 & 0 \\ 
 30 & FGC 0495 & 4430.0 & Scd & 405.3 & 16.3 & 15.7 & 9.4 & 9.4 & -0.5 & 0 \\ 
 30 & MCG -01-12-037 & 4797.0 & Sc & 241.2 & 15.4 & 14.9 & 9.9 & 8.6 & 0.3 & 0 \\ 
 30 & W043547.08-033114.6 & 4688.0 & Sm & 737.9 & 18.4 & 18.0 & 8.4 & 9.0 & -0.4 & 0 \\ 
 30 & W043404.12-032444.0 & 4588.0 & Sd & 865.8 & 17.4 & 17.0 & 8.8 & 8.5 & 0.5 & 0 \\ 
 31 & W045913.48-033631.4 & 3630.0 & Sm & 824.6 & 15.9 & 15.6 & 9.2 & 8.7 & 0.6 & 0 \\ 
 31 & W050042.68-035434.1 & 4033.0 & E & 392.5 & 18.8 & 17.9 & 9.1 & 8.6 & 0.7 & 0 \\ 
 31 & W050300.01-042846.1 & 4002.0 & Sm & 369.1 & 16.7 & 16.5 & 8.7 & 9.3 & -0.3 & 0 \\ 
 31 & W050112.74-040732.1 & 4333.0 & Sc & 162.9 & 17.0 & 16.5 & 9.0 & 8.7 & 0.5 & 0 \\ 
 31 & W045925.54-035506.3 & 4070.0 & Sd & 604.6 & 17.1 & 16.7 & 8.7 & 8.9 & 0.0 & 0 \\ 
 31 & W050152.03-035116.1 & 4006.0 & S0 & 380.9 & 17.9 & 17.5 & 8.4 & 7.8 & 1.0 & 0 \\ 
 31 & G050203.46-040132.9 & 4091.0 & Im & 238.0 & 19.3 & 19.0 & 7.8 & 7.7 & 0.3 & 0 \\ 
 90 & J22012971-3157464 & 2785.0 & S0 & 72.0 & 16.7 & 16.1 & 8.8 & 7.6 & 1.6 & 0 \\ 
 90 & ESO 466- G 036 & 2379.0 & Sa & 251.9 & 14.4 & 13.8 & 9.7 & 9.4 & -0.8 & 0 \\ 
 90 & W220250.16-323436.9 & 2200.0 & S0 & 379.3 & 16.3 & 15.8 & 8.9 & 8.1 & 1.1 & 0 \\ 
 90 & W220521.01-320000.2 & 2256.0 & Sb & 401.8 & 15.4 & 14.9 & 9.3 & 8.1 & 0.5 & 0 \\ 
 90 & NGC 7163 & 2754.0 & SBab & 335.6 & 13.6 & 12.9 & 10.3 & 8.4 & 0.5 & 0 \\ 
 90 & B215808.24-321139.7 & 2642.0 & S0 & 125.4 & 17.9 & 17.6 & 7.9 & 7.8 & 0.4 & 0 \\ 
 90 & ESO 466- G 046 & 2318.0 & S0a & 86.3 & 14.8 & 14.0 & 9.8 & 7.0 & 1.5 & 1 \\ 
 90 & ESO 466- G 047 & 2556.0 & Sbc & 85.3 & 16.1 & 15.5 & 9.1 & 7.0 & 2.3 & 1 \\ 
 90 & ESO 466- G 044 & 2818.0 & S0 & 107.3 & 15.2 & 14.5 & 9.6 & 7.0 & 1.3 & 1 \\ 
 90 & ESO 466- G 051 & 2632.0 & S0 & 211.8 & 14.2 & 13.5 & 10.0 & 7.1 & 1.6 & 1 \\ 
 90 & S408Z037 & 2546.0 & SBc & 292.3 & 13.7 & 13.1 & 10.0 & 7.2 & 1.9 & 1 \\ 
 90 & ESO 404- G 028 & 2339.0 & Sa & 466.3 & 14.8 & 14.1 & 9.8 & 7.7 & 0.7 & 1 \\ 
 91 & W220943.70-273554.8 & 7072.0 & S0 & 372.3 & 18.7 & 18.1 & 8.9 & 8.1 & 1.1 & 0 \\ 
 91 & ESO 467- G 016 & 7372.0 & Sa & 479.0 & 14.9 & 14.2 & 10.5 & 9.6 & -0.3 & 0 \\ 
 91 & ESO 467- G 014 & 7112.5 & SBc & 627.3 & 15.9 & 15.7 & 9.3 & 9.8 & -0.8 & 0 \\ 
 91 & ESO 467- G 003 & 6975.0 & SBbc & 1068.2 & -- & -- & -- & 9.9 & -- & 0 \\ 
 91 & ESO 467- G 005 & 6868.5 & Sb & 974.0 & 15.5 & 15.0 & 10.0 & 9.6 & -0.4 & 0 \\ 
 91 & ESO 532- G 023 & 7235.0 & SBbc & 893.4 & 15.2 & 14.7 & 10.2 & 9.8 & -0.5 & 0 \\ 
 91 & W221035.73-271456.0 & 7180.0 & S0 & 1012.7 & 17.0 & 16.6 & 9.2 & 8.6 & -0.2 & 0 \\ 
 91 & W221040.40-274144.2 & 6814.0 & Sd & 560.1 & 18.4 & 17.9 & 8.7 & 9.1 & -0.0 & 0 \\ 
 91 & W220659.01-271620.4 & 6931.0 & Im & 1140.7 & -- & -- & -- & 9.0 & -- & 0 \\ 
 91 & W220836.49-281651.5 & 6966.0 & Im & 806.6 & -- & -- & -- & 9.0 & -- & 0 \\ 
 91 & W220813.38-272252.5 & 7265.0 & S0 & 740.6 & -- & -- & -- & 8.5 & -- & 0 \\ 
 91 & W220913.03-273402.8 & 6990.5 & E & 363.6 & 15.5 & 14.8 & 10.4 & 7.9 & 1.2 & 1 \\ 
 91 & W220938.70-273316.5 & 7435.0 & E & 420.1 & 18.9 & 18.5 & 8.4 & 7.9 & 0.8 & 1 \\ 
 91 & W220727.04-275534.3 & 6923.0 & S0 & 643.0 & -- & -- & -- & 8.0 & -- & 1 \\ 
 91 & W221011.26-281122.4 & 7016.0 & S0 & 732.8 & 15.5 & 14.8 & 10.4 & 8.1 & 1.0 & 1 \\ 
 97 & W234720.25-022225.9 & 7247.0 & S0a & 94.8 & 17.5 & 17.3 & 8.7 & 8.6 & 0.5 & 0 \\ 
 97 & LEDA 1092309 & 7235.0 & Sa & 223.9 & 17.1 & 16.6 & 9.3 & 8.8 & -0.5 & 0 \\ 
 97 & NSA 152646 & 6457.0 & S0 & 211.8 & 17.8 & 17.3 & 9.0 & 8.7 & 0.5 & 0 \\ 
 97 & NSA 152613 & 6701.0 & Sd & 259.3 & 18.0 & 17.6 & 8.9 & 8.5 & 0.7 & 0 \\ 
 97 & W234840.59-022144.1 & 6674.0 & Sab & 479.6 & 15.8 & 15.2 & 10.0 & 8.8 & 0.0 & 0 \\ 
 97 & W234920.76-022329.2 & 7434.0 & Sa & 730.6 & 17.5 & 17.0 & 9.1 & 8.6 & 0.7 & 0 \\ 
 97 & W234528.37-021158.5 & 6609.0 & Sbc & 732.9 & 16.4 & 15.9 & 9.6 & 9.0 & -0.2 & 0 \\ 
 97 & W234759.93-023431.4 & 6974.5 & Sa & 449.7 & 15.7 & 15.2 & 10.0 & 9.1 & -0.3 & 0 \\ 
 97 & W234731.71-022524.9 & 6223.0 & Sm & 172.3 & 17.5 & 17.1 & 9.0 & 8.8 & 0.4 & 0 \\ 
 97 & UM 177 & 6505.0 & Sc & 459.8 & 17.3 & 17.0 & 8.9 & 9.2 & -0.0 & 0 \\ 
 97 & W234741.96-022952.2 & 6491.0 & Sbc & 297.5 & 18.3 & 17.9 & 8.7 & 8.4 & 0.6 & 0 \\ 
 97 & W234606.83-021248.9 & 6626.0 & Sm & 498.1 & 17.7 & 17.3 & 8.8 & 8.9 & 0.2 & 0 \\ 
 97 & W234701.65-022033.4 & 6767.0 & Im & 145.9 & 17.4 & 17.0 & 9.2 & 9.1 & 0.1 & 0 \\ 
 97 & W234446.64-020140.0 & 6821.0 & Sa & 1059.1 & 17.1 & 16.7 & 9.2 & 8.7 & -0.4 & 0 \\ 
 97 & W234808.62-023634.0 & 6949.0 & Sab & 520.5 & 17.8 & 17.3 & 9.0 & 9.1 & 0.1 & 0 \\ 
 97 & W234646.78-015151.0 & 6926.0 & Sd & 702.9 & 17.6 & 17.2 & 9.0 & 8.8 & 0.5 & 0 \\ 
 97 & W234714.35-015813.3 & 6947.0 & Sm & 511.0 & 16.2 & 15.8 & 9.6 & 8.6 & 0.4 & 0 \\ 
 97 & G234617.06-015255.0 & 7012.0 & Im & 759.9 & 19.4 & 19.0 & 8.2 & 8.4 & 0.1 & 0 \\ 
 97 & W234737.41-022422.4 & 7051.0 & Sm & 162.6 & 18.1 & 17.8 & 8.6 & 8.9 & -0.0 & 0 \\ 
 97 & W234705.08-023423.8 & 7194.0 & Scd & 404.8 & 18.2 & 17.8 & 8.7 & 8.9 & 0.1 & 0 \\ 
 97 & G234535.72-021211.7 & 7214.0 & S0 & 687.6 & 17.6 & 17.1 & 9.1 & 8.6 & 0.6 & 0 \\ 
 97 & W234603.39-021824.5 & 7171.0 & Sab & 497.7 & 17.4 & 17.1 & 8.8 & 9.2 & -0.0 & 0 \\ 
 97 & W234731.08-023340.4 & 7320.0 & S0 & 372.4 & 18.0 & 17.6 & 8.7 & 8.4 & 0.6 & 0 \\ 
 97 & W234750.99-023053.6 & 7364.0 & Sd & 344.5 & 18.3 & 17.9 & 8.7 & 8.5 & 0.4 & 0 \\ 
 97 & MCG -01-60-040 & 6379.0 & Sb & 288.3 & 15.1 & 14.6 & 10.3 & -- & -- & 0 \\ 
 97 & MCG -01-60-041 & 6682.0 & Sa & 343.0 & 14.0 & 0.3 & 10.9 & -- & -- & 0 \\ 
 97 & W234732.41-022229.7 & 6879.0 & E & 107.0 & 19.2 & 18.5 & 8.9 & 7.7 & 1.4 & 1 \\ 
 97 & W234659.42-020811.6 & 6459.0 & S0 & 301.1 & 17.1 & 16.4 & 9.7 & 7.8 & 0.6 & 1 \\ 
 97 & X23464481-0210169 & 6894.0 & E & 318.3 & 15.9 & 15.0 & 10.5 & 7.8 & 1.1 & 1 \\ 
 97 & 2MFGC 17840 & 6447.0 & S0 & 415.0 & 16.9 & 16.1 & 9.8 & 7.8 & 0.7 & 1 \\ 
 97 & W234718.75-015943.3 & 6853.0 & E & 470.8 & 16.0 & 15.3 & 10.2 & 7.8 & 1.0 & 1 \\ 
 \hline
\end{longtable}}